\newif\ifsubmode
\submodefalse

\newif\ifprintfig
\printfigtrue

\ifsubmode
 \documentclass[epsf,manuscript]{aastex}
  \received{}
  \accepted{}
  \journalid{}{}
  \articleid{}{}
  \slugcomment{Submitted to {\it PASP}}
  \shorttitle{Telluric Corrections in the Near-Infrared}
  \shortauthors{Vacca et al.}
\else
 \documentclass[12pt,preprint,epsf]{aastex}
  \shorttitle{Telluric Corrections in the Near-Infrared}
  \shortauthors{Vacca et al.}
  \slugcomment{Accepted for publication in {\it PASP}}
\fi
 
\lefthead{Vacca, Cushing, Rayner}
\righthead{Telluric Corrections in the Near-Infrared}

\def\bdm{\begin{displaymath}} 
\def\edm{\end{displaymath}} 
\def\beq{\begin{equation}} 
\def\eeq{\end{equation}} 
\def\bit{\begin{itemize}} 
\def\eit{\end{itemize}} 
\def\ben{\begin{enumerate}} 
\def\een{\end{enumerate}} 
\def\bfi{\begin{figure}[htb]} 
\def\bpfi{\begin{figure}[p]}
\def\mm{$\rm \mu m$}

\def\epsfbox#1{}
 
\begin{document}

\title{A Method of Correcting Near-Infrared Spectra for
Telluric Absorption\altaffilmark{1}}

\author{William D.\ Vacca\altaffilmark{2}, Michael C. Cushing, and John T.\ Rayner} 
\affil{Institute for Astronomy, Honolulu, HI 96822}
\authoremail{vacca@mpe.mpg.de;cushing,rayner@ifa.hawaii.edu}

%%%%%%%%%%%%%%
% Additional affiliations
%%%%%%%%%%%%%%

\altaffiltext{1}{Based on observations obtained with the Infrared Telescope 
Facility, which is operated by the University of Hawaii under contract
to the National Aeronautics and Space Administration.}
\altaffiltext{2}{Currently at the Max-Planck-Institut f\"ur Extraterrestrische 
Physik, Postfach 1312, Giessenbachstrasse, 85741 Garching, Germany}

%%%%%%%%%%%%%%%
% Start the abstract on a fresh page
%%%%%%%%%%%%%%%
 
\ifsubmode\else
\clearpage\fi

%%%%%%%%%%%%%%%
% Use a small baselineskip, unless in submission mode.
%%%%%%%%%%%%%%%
 
\ifsubmode\else
\baselineskip=14pt
\fi

%%%%%%%%%%%%%%%
% Abstract
%%%%%%%%%%%%%%%

\begin{abstract}
 
We present a method for correcting near-infrared medium-resolution
spectra for telluric absorption. The method makes use of a spectrum 
of an A0V star, observed near in time and close in airmass to the target object, 
and a high-resolution model of Vega, to construct a telluric
correction spectrum that is free of stellar absorption features. The
technique was designed specifically to perform telluric corrections on 
spectra obtained with SpeX, a $0.8-5.5$\mm ~medium-resolution cross-dispersed 
spectrograph at the NASA Infrared Telescope Facility, and uses the fact 
that for medium resolutions there exist spectral regions
uncontaminated by atmospheric absorption lines.
However, it is also applicable
(in a somewhat modified form) to spectra obtained with other near-infrared
spectrographs. An IDL-based code that carries out the procedures is 
available for downloading via the World Wide Web. 

\end{abstract}

%%%%%%%%%%%%%%%
% Keywords
%%%%%%%%%%%%%%%

\keywords{techniques: spectroscopic --- methods: data analysis --- atmospheric effects}
          
\clearpage

%%%%%%%%%%%%%%%
% Beginning of main text
%%%%%%%%%%%%%%%

\section{Introduction}
\label{s:intro}

Ground-based near-infrared spectroscopy ($\sim 1-5$ \mm) has always been hampered 
by the strong and variable absorption features arising from the Earth's atmosphere.
Even within the well-established photometric bands such as $J$, $H$, and $K$,
so-called ``telluric absorption features" are present. The common method used 
to correct for such absorption features, and to retrieve the signal of the
object under observation, is to observe a ``telluric standard star", near in both
time and sky position (airmass) to the object, and subsequently to divide the object's
spectrum by the standard star's spectrum during the final phases of the reduction
process. 

A-type stars are frequently used as such telluric standards, as their
spectra contain relatively few metal lines, with strengths that are generally only a 
few percent of the continuum.
Furthermore, for A stars with little reddening, the magnitudes
at all wavelengths can be deduced from a single (usually $V$-band) measurement.
Finally, the near-infrared continua of A stars can be reasonably well approximated by a 
blackbody with a temperature of $\sim 10000~ K$. These latter two properties make
A-type stars very useful for determining the absolute flux calibration of the observations.
However, the strong intrinsic photospheric hydrogen (H) absorption features in 
the spectra of A-type stars present a substantial
problem for their use as telluric standards, particularly if one is 
working in the $I$, $z$, $Y$, or $H$ bands (where the numerous higher level lines of the 
Paschen and Brackett series are located), or if the observations are designed to
study the H lines in the spectrum of the object of interest. One technique 
commonly adopted in order to preserve the other desirable qualities of A stars
as telluric and flux standards is to simply interpolate over the H absorption lines in 
their observed spectra. However, since
atmospheric absorption features are certainly present at the locations of most of these lines,
this method can lead to uncertainties in the reality of any features subsequently
detected in the corrected object's spectrum at or near the wavelengths of the stellar H lines.
Furthermore, A-type stars can have rather large rotational velocities (upwards of
300 km s$^{-1}$; e.g., Abt \& Morrelli 1995; Royer et al.\ 2002), which broaden the
absorption lines. The large number of such broad lines located near the various H
series limits renders the interpolation method infeasible.

Maiolino, Rieke, \& Rieke (1996) developed a correction method using G2V stars,
rather than A-type stars, as telluric standards. Their method makes use of
the high signal-to-noise (S/N), high-resolution spectrum of the Sun, obtained
using a differential airmass technique
at the McMath Telescope at Kitt Peak National Observatory by Livingston \& Wallace (1991; 
see also Wallace, Hinkle, \& Livingston 1993), to generate a telluric correction spectrum
which is then applied to the spectrum of the object of interest. The
correction spectrum is derived from the ratio of the solar spectrum
(shifted to the appropriate radial velocity and convolved to the
resolution of the observations) to the observed G2V star spectrum. This
ratio should, in principle, remove any intrinsic stellar features from
the telluric correction spectrum. Although Maiolino et al.\ (1996)
demonstrated that this method can be reasonably successful in performing
telluric corrections on moderate S/N and resolution data, there are
several practical problems with it, especially when applying it to higher
quality and resolution spectra. Firstly, the number of reasonably
bright G2V stars available for use as telluric standards near any given
object is less than the number of A-type stars, simply because
G stars are inherently less luminous than A stars and catalogues of stars with
well-defined spectral classifications are based on magnitude-limited samples.
This is demonstrated in Table 1, in which we have compiled the numbers of
stars listed with A0V and G2V spectral types in the SIMBAD database as a
function of $V$ magnitude. The intrinsic $(V-K)$ color of a G2V star is
about $1.5$ mag; therefore, even in the $K$ band, bright
A0V stars outnumber G2V stars in the database.
Secondly, G-type stars have a large number of (relatively weak) metal lines throughout their
spectra. Hence the accuracy of the resulting corrections depends on how well
the observed spectrum of the G star matches that of the Sun; the higher the 
S/N and resolution of the data, the closer the spectral match must
be for a desired level of accuracy. As the strengths of
the metal lines vary considerably with spectral type and stellar parameters, any mismatch, such
as might be caused by spectral classification inaccuracies, metal abundance
differences, anomalous line strengths, etc., will introduce spurious features
into the telluric correction spectrum. Thirdly, the reference solar spectrum was obtained 
at a relatively low site and has large wavelength gaps where the intrinsic spectrum could 
not be recovered using the differential airmass method. 
Because the depths of atmospheric absorption features vary with elevation, 
this solar spectrum would not provide a useful correction
template for spectra, such as those obtained at a high site such as Mauna Kea, which have 
recoverable flux in the regions of strong absorption between the photometric passbands.

An alternative method, used for example by Hanson, Conti, \& Rieke (1996), combines the 
A-star technique with that of Maiolino et al.\ (1996) and involves the following steps: 
observe both A-type stars and G2V stars, normalize the spectra of both
stars, ``correct'' the normalized G2V star spectrum by dividing it by the solar spectrum,
replace the regions centered on the H lines in the normalized A star spectrum
by the corrected G star spectrum, and finally divide the result into
the object's spectrum. This is equivalent to dividing the object
spectrum by that of the A star, and multiplying the result by a line correction
spectrum generated from the G2V star and the solar spectrum. Since the solar spectrum
is used only in the regions of the H lines, the resulting correction template
should be more accurate than one based on the G2V star/solar spectrum ratio alone.
Because the normalization
process removes the signature of the instrumental throughput from the observed
stellar spectra, the object must be flux-calibrated separately, after the correction
process, usually by multiplying the normalized, corrected
spectrum by a blackbody with a temperature appropriate for an A star. While this method might
seem attractive in principle, in practice it requires observations of two
sets of standards (a suitable one of which may be difficult to find, as explained above) 
for each object of interest (ideally)
and involves a somewhat tedious and complicated reduction 
procedure, particularly if the wavelength range of interest includes several absorption
lines of the Pa or Br series. 
%It also suffers from the difficulty of finding
%appropriate G2V stars near the target object.

Clearly, the ideal telluric standard would be one whose spectrum is completely
featureless and precisely known. In addition, such objects should be reasonably bright,
fairly common, and widely distributed on the sky, so that one can be found close to any 
particular object of interest. While few objects satisfying the first criterion
are known in nature, there are a limited number of objects for which precise and accurate
theoretical models are available. Vega is one such object. In this paper we describe
a new method of generating a telluric correction spectrum using observations of
A0V stars and a high-resolution model of Vega's intrinsic spectrum. Through a process
somewhat similar to that described by Maiolino et al.\ (1996), this technique
preserves the desirable qualities of A-type stars, while avoiding the problems with the
intrinsic stellar absorption lines, and does not require observations of G2V stars.

The method outlined below was developed for the reduction of spectra
obtained with SpeX at the NASA Infrared Telescope Facility (IRTF) on
Mauna Kea. SpeX is a medium resolution ($\lambda/\Delta\lambda \sim 200-2500$)
cross-dispersed spectrograph equipped with a $1024\times 1024$ InSb array which
provides simultaneous wavelength coverage over $0.8-2.4~$\mm\ in one
grating setting and $2.4-5.5~$\mm\ in another grating setting (see
Rayner et al.\ 1998, 2002 for a full description of SpeX and its
various scientific observing modes). 
Because the entire Paschen series and most of the Brackett series of H absorption lines
are simultaneously detected in a spectrum of any A0V star obtained with SpeX,
it was important to devise a technique that could be applied easily and rapidly to 
accurately remove all of these features from a telluric correction spectrum
generated from the A0V star spectrum. With high quality spectra of an A0V star, a 
telluric correction spectrum accurate to better than about 2 percent across the H lines 
can be quickly constructed with this technique. Although designed specifically for SpeX
data, the method can also be used to correct spectra obtained with most medium resolution
($\lambda/\Delta\lambda < 50000$) near-infrared spectrographs. 
A set of IDL-based routines (Spextool\_extension, v.\ 2.0), specifically designed 
for application to SpeX data, implements the telluric correction method described below and 
comprises an extension to Spextool (v.\ 2.3), a package of routines written for the
basic reduction of spectra obtained with SpeX (Cushing, Vacca, \& Rayner 2002). Both sets 
of programs are available from the SpeX webpage at the IRTF 
website ({\tt http://irtfweb.ifa.hawaii.edu/Facility/spex}). Finally, a third set 
of IDL routines, constituting a general version 
(i.e., not specific to SpeX data) of the telluric correction code, is also
available from this webpage.

\section{Description of Method}
\label{s:method}

We assume that the observed spectrum $O(\lambda)$ of any object is given by
\begin{equation}
O(\lambda) = {[I(\lambda) \cdot T(\lambda)] \ast P(\lambda)} \cdot Q(\lambda),
\label{basiceq}
\end{equation}
where $I(\lambda)$ is the intrinsic spectrum of the object, $T(\lambda)$ is the
atmospheric (telluric) absorption spectrum, $P(\lambda)$ is the instrumental profile,
$Q(\lambda)$ is the instrumental throughput, and $\ast$ denotes a convolution.
Because the convolution represents a smoothing of the intrinsic stellar and telluric
spectra, the above equation can be re-written as
\begin{equation}
O(\lambda) = {[I(\lambda) \ast P(\lambda)] \cdot [T(\lambda) \ast P(\lambda)]} \cdot Q(\lambda),
\label{basiceq2}
\end{equation}
or 
\begin{equation}
O(\lambda) = I_{instr}(\lambda) \cdot R(\lambda),
\label{basiceq3}
\end{equation}
where 
\begin{equation}
I_{instr}(\lambda) = I(\lambda) \ast P(\lambda)
\label{Idefn} 
\end{equation}
is the spectrum of the 
object observed above the atmosphere with an instrument with perfect throughput, 
and 
\begin{equation}
R(\lambda) = [T(\lambda) \ast P(\lambda)] \cdot Q(\lambda) 
\end{equation}
is the telluric
absorption spectrum convolved with the instrument profile and multiplied by the 
instrumental throughput. The latter represents the overall throughput,
or response, of the atmosphere$+$instrument system. We assume that $Q(\lambda)$ is 
a smooth, featureless function and therefore any observed high frequency 
variations in $O(\lambda)$ must be due to either $I(\lambda)$ or $T(\lambda)$. 

The immediate objective is to determine $I_{instr}$ of a specific object  
from $O(\lambda)/R(\lambda)$.
Conversely, if $I(\lambda)$ for any particular source is known {\it a priori}, 
and $P(\lambda)$ of the instrument is known or can be determined, 
then the system throughput $R(\lambda)$ can be determined from Eqs.\ (\ref{basiceq3})
and (\ref{Idefn}).
This factor can then be used to determine $I_{instr}(\lambda)$ of a target
object from the observed spectrum, provided that the telluric absorption spectrum 
$T(\lambda)$ does not vary substantially between the two sets
of observations. This forms the basis of both our method and that of
Maiolino et al.\ (1996): observations of a ``telluric standard'', whose intrinsic spectrum is
known, to derive a system throughput curve which can then be applied to a target object spectrum
observed nearby in the sky and close in time. Since $T(\lambda)$ varies with airmass,
and on time scales of the order of several to tens of minutes (depending on the atmospheric
conditions), it is clearly best to observe a 
telluric standard at an airmass as close as possible to that of the target object, and
within a few minutes of the observations of the target.

Because the metal lines in A0V spectra are relatively infrequent and weak (on the order of 
a few percent of the continuum level in the near-infrared at the resolution of SpeX; 
see, for example, Fig.\ 1), 
the intrinsic near-infrared spectra of such objects consist essentially of H absorption 
lines and a smooth, featureless (blackbody) continuum. Thus, metal abundance variations
do not strongly affect the appearance of the observed spectra and, except for 
possible differences in the widths and 
depths of some of the H lines (due to different rotational velocities or slightly different
values of the surface gravity), the spectra of all A0V stars can be considered nearly
identical to one another. It is a fortunate circumstance that Vega, the
object that defines the zeropoint of the magnitude system, is also a
bright A0V star whose spectrum has been accurately modeled. Assuming the model provides 
an accurate representation of the intrinsic spectrum of Vega, and that Vega is a good
representative (an archetypal example) of the class of A0V stars, we can use 
the theoretical model to determine $I_{instr}(\lambda)$ of any A0V star.
Hence, if each observation of a given target object is preceded and/or followed by an
observation of an A0V star, and if the model spectrum of Vega can be scaled and reddened 
to match the near-infrared magnitudes of the observed A0V star, and modified to
account for the differences in line strengths, radial and rotational
velocities, and spectral resolution, we will be in the happy situation of having
at hand both the observed spectrum, $O(\lambda)$, and intrinsic spectra, $I(\lambda)$
and $I_{instr}(\lambda)$, 
%(if $P(\lambda)$ is known)
of the A0V star. The system response function, $R(\lambda)$, 
% or ``telluric correction spectrum'', 
can then be determined immediately and used to correct the target object's spectrum. 
%Now, suppose that each observation of any given target object is preceded and/or
%followed by an observation of A0V star, near in time and close in airmass to those of the
%object of interest. The observed spectrum of the A0V star and the
%modified model of Vega can be used to construct a ``telluric correction
%spectrum'', which can then be used to correct the target object's spectrum. 
The difficulty then lies ``only'' in properly modifying the model spectrum of Vega.

Modifying the model Vega spectrum involves shifting it to the radial
velocity of the observed A0V star, scaling and reddening it to reproduce the known 
magnitudes of the observed A0V star, convolving it with a function (kernel) that
broadens the lines to the observed widths and smooths it to the
observed resolution, resampling the result to the observed wavelength scale, and 
altering the depths of
various H lines to match those of the observed A0V star. The convolution kernel 
is given by the instrument profile (IP), $P(\lambda)$, and a (rotational) broadening
function $\Theta (\lambda)$ (see below). There are at least three ways 
of constructing $P(\lambda)$: (1) assume a theoretical profile whose properties are 
defined by the instrument parameters (e.g., slit width); (2) construct an empirical 
profile using observations of
unresolved arc lines (similar to those used to perform the wavelength
calibration of the spectra); and (3) construct a profile using the
observations of the A0V star itself. We discuss the last method in
detail here, and describe below a set of routines that implement this method for
use in the reduction of data obtained with SpeX. (The routines also allow
users to adopt empirical or theoretical instrument profiles.)

We begin by selecting a wavelength region centered on an intrinsic absorption line in the
observed and model A0V spectra. The observed spectrum is normalized 
in the region of the line by fitting a low order polynomial or a spline
to the continuum; a good representation of the observed continuum can be obtained 
once bad pixels and the
regions around any strong absorption features are excluded from the fitting process. The
normalization removes the signature of the instrumental throughput as
well as the shape of the stellar continuum. Once the observed stellar spectrum 
has been normalized, the wavelength region around the absorption feature can be
cross-correlated with that from the normalized model of Vega, using a technique
similar to that described by Tonry \& Davis (1979), to determine the
observed radial velocity of the A0V star. The model is then shifted by
the estimated radial velocity, and scaled and reddened to match the observed magnitudes of
the A0V star.

We now wish to construct a convolution kernel which, when applied to
the model spectrum of Vega, will produce a close approximation of the
intrinsic spectrum of the observed A0V star, as measured by an instrument
with perfect throughput. That is, we assume that the intrinsic spectrum of the 
A0V star can be given by
\begin{equation}
I_{A0V}(\lambda) = I_{Vega}(\lambda) \ast \Theta(\lambda)
\end{equation}
where $I_{Vega}(\lambda)$ is the scaled and shifted model spectrum of Vega and
$\Theta(\lambda)$ is a function that accounts for the differences in
rotational velocity and any other line broadening factors between the Vega model
and the observed A0V star. The observed A0V spectrum $O_{A0V}(\lambda)$ is then given by
\begin{equation}
O_{A0V}(\lambda) = \{[I_{Vega}(\lambda) \ast \Theta(\lambda)] \cdot T(\lambda)\} 
\ast P(\lambda) \cdot Q(\lambda),
\label{conveq}
\end{equation}
or,
\begin{equation}
O_{A0V}(\lambda) = [I_{Vega}(\lambda) \ast K(\lambda)] \cdot R(\lambda),
\label{Keq}
\end{equation}
where 
\begin{equation}
K(\lambda) = \Theta(\lambda) \ast P(\lambda),
\label{Kdefn}
\end{equation} 
and 
\begin{equation}
I_{instr,A0V} = I_{Vega}(\lambda) \ast K(\lambda).
\end{equation}
For the moment, let us simply assume we can find an intrinsic stellar absorption line in 
the observed A0V spectrum across which both the atmospheric absorption and the instrumental
throughput are either constant or only smoothly varying. Using the normalized
observed and model spectra in the vicinity of the line, we then have,
\begin{equation}
O'_{A0V}(\lambda) = I'_{Vega}(\lambda) \ast K(\lambda),
\end{equation}
where primes denote the normalized spectra (divided by the respective continua).
The convolution kernel $K(\lambda)$ can then be determined by selecting a small region around
the observed absorption line and performing a deconvolution.

Once $K(\lambda)$ has been determined, it is convolved with the shifted, scaled, and reddened 
Vega model spectrum and
%that has been scaled and reddened to match the observed magnitudes of the A0V star, and 
$R(\lambda)$ is determined from equation (\ref{Keq}). Under the assumptions that all A0V stars 
are identical to Vega and that the model of Vega provides a highly accurate representation of
Vega's intrinsic spectrum, the convolution and subsequent division,
$O_{A0V}(\lambda) / [I_{Vega}(\lambda) \ast K(\lambda)]$ should yield a response 
spectrum in which the intrinsic stellar absorption lines are absent. In practice, we find that
some small residual features at the locations of the H lines are discernible in this response 
spectrum due to differences in the strengths of these lines between the model and the observed 
A0V spectra. Some of these differences may in fact be real, due to different values of $\log g$ between
the standard star and the Vega or misclassification of the observed
standard star and hence a slight spectral mismatch between the standard and Vega.
However, the model H line strengths (equivalent widths) can be easily adjusted to minimize, 
or eliminate, these residuals. Furthermore, since any metal lines are fairly weak and
do not vary strongly among late B-type and early A-type stars, the effects of spectral mismatches and 
misclassifications 
%of stars used as standards 
on the resulting telluric correction spectrum
can be minimized to a large extent by simply modifying the H line strengths in the model. For 
the resolving power and typical S/N levels achieved with SpeX, minor misclassifications
should have a negligible effect on the final corrected spectra. 
 
The intrinsic spectrum of the target object $I_{instr}(\lambda)$ is then determined from 
equation (\ref{basiceq3}), re-written as
\begin{equation}
I_{instr}(\lambda) = O(\lambda) \cdot [I_{Vega}(\lambda) \ast K(\lambda) / O_{A0V}(\lambda)].
\end{equation}
We shall refer to the quantity $[I_{Vega}(\lambda) \ast K(\lambda)] / O_{A0V}(\lambda) = 1/R(\lambda)$ 
as the ``telluric correction spectrum''.
One advantage of this method is that the resulting spectrum $I_{instr}(\lambda)$ is automatically 
placed on the proper flux scale, if the slit losses during the two sets of observations (standard
star and target object) were the same.  

%This method assumes that $P(\lambda)$ does not vary substantially across the spectrum.
%Most spectrographs are designed such that the spectral resolution ($\lambda/\Delta\lambda$)
%is approximately constant across the observed wavelength range. In this case, $P(\lambda)$
%is not constant; however, the instrument profile will be constant in velocity, or $\log \lambda$,
%or pixel coordinates instead. All of the above equations can then be carried out in these
%coordinates. Our own experience has demonstrated that variations in the spectral
%resolution of $\pm 10$\% have minimal effect on the quality of the results.

In summary, the method of constructing a telluric correction spectrum 
consists of an observation of an A0V star and several steps designed to modify the model spectrum 
of Vega appropriately:
\begin{enumerate}
  \item Normalization of the observed A0V star spectrum in the vicinity of a suitable
        absorption feature (as defined below);
  \item Determination of the radial velocity shift of the A0V star;
  \item Shifting the Vega model spectrum to the radial velocity of the A0V star;
  \item Scaling and reddening the Vega model spectrum to match the observed magnitudes of the A0V star;
  \item Construction of a convolution kernel from a small region around an absorption feature in the 
        normalized observed A0V and model Vega spectra;
  \item Convolution of the kernel with the shifted, scaled, and reddened model of Vega;
  \item Scaling the equivalent widths of the various H lines to match those of the observed A0V star.
\end{enumerate}
Finally, the convolved model is divided by the observed A0V spectrum and the resulting telluric
correction spectrum is multiplied by the observed target spectrum.

\section{Practical Considerations}
\label{s:practical}

For the method described above to work properly, the convolution kernel
$K(\lambda)$ must be determined accurately. If $P(\lambda)$ and $\Theta(\lambda)$ are known in
advance, $K(\lambda)$ can be constructed using equation \ref{Kdefn}. If $K(\lambda)$ is to be
generated from the observations themselves, it is necessary to find a
region of the observed A0V spectrum which contains a strong intrinsic stellar
absorption line (so that the radial and rotational velocities and IP can be
determined), but is free from atmospheric absorption or instrumental features.
The latter requirement arises from our assumption that $T(\lambda)$ 
and $Q(\lambda)$ are either constant or vary only slowly and smoothly across the width of the line.
Fortunately, for observations with SpeX, 
the atmosphere has cooperated in providing such a wavelength region and
stellar atmospheres have cooperated in providing a suitable absorption line.
Fig.\ \ref{f:absorption} presents a model of the atmospheric
transmission between 0.9 and 1.2 \mm, computed with the ATRAN program (Lord 1992) for an
airmass of 1.2 and standard conditions on Mauna Kea and smoothed to a resolving power
of 2000. The normalized
model spectrum of Vega obtained from R.\ Kurucz ({\tt http://kurucz.harvard.edu/stars.html}), 
also smoothed to resolving power of 2000, is overplotted. 
It can be seen that the Pa $\delta$ (1.00494 \mm) absorption line in the Vega spectrum 
is located in a region where, at this resolution, atmospheric absorption is almost completely
negligible. Therefore, this line can be used to determine the
convolution kernel $K(\lambda)$ in the procedure described above.

In addition to the lack of telluric contamination, the Pa $\delta$ line
also lies in a wavelength region where the throughput $Q(\lambda)$ of SpeX is 
relatively high and does not change dramatically across the width of the line. 
The throughput around the Pa $\gamma$
(1.0938 \mm) line, for example, drops quickly across the line (due to the grating
blaze function) and atmospheric absorption affects the shape of
the long wavelength wing, making the
determination of the continuum near the wings difficult. Nevertheless,
although the Pa $\delta$ line is ideal for our purposes,
we have successfully generated telluric correction spectra using the method described
in Section 2 with kernels built from other H lines in the spectra of A0V stars
(e.g., Br $\gamma$). For spectrographs that do not contain cross-dispersers and/or 
yield only a limited wavelength range, the observed A0V spectra may not contain a line
that is suitable for constructing $K(\lambda)$. In these cases,
the convolution kernel can be generated from the IP and some estimate of the rotational
velocity of the A0V star (Eq.\ \ref{Kdefn}). At resolving powers similar to those of 
SpeX, the latter generally has a negligible effect on the resulting telluric correction
spectrum.

The model of Vega we use for our telluric correction procedure has a resolving power 
($\lambda / \Delta\lambda$) of 500000, binned by 5 (giving an effective resolving 
power of 100000), and has been scaled to match the observed Vega flux at 5556 \AA\ 
given by M$\rm \acute{e}$gessier (1995). (A higher resolution model
is also available from the Kurucz web page; however the array lengths necessary to
store these data and the subsequent intermediate products become
inordinately large and unwieldy.) The properties of the model are as follows:
$T_{\rm eff} = 9550~$K, $\log g = 3.950$, $v_{\rm rot} = 25~$ km/s,  
$v_{\rm turb} = 2~$ km/s, and a metal abundance somewhat below solar. The  
model also has an associated continuum spectrum; using the full spectrum and
the continuum, we generated a normalized 
line spectrum. The continuum spectrum contains sharp jumps due to the abrupt
increases/decreases in the opacity at the limits of the H (Pa and Br) series. The
strengths of these jumps must be modified along with the line strengths.
To account for this, we have normalized the continuum spectrum by dividing
it by a spline fitted to the spectrum far from the jumps. The fit passes approximately
through the flux midpoints between the continuum levels on either side of the jumps. Scaling 
this normalized
continuum by an appropriate value then has the effect of scaling the heights
of the jumps. 

Note that because the Vega model spectrum has such a large resolving
power, the method we have outlined should work successfully for spectral data with considerably larger 
resolving powers than that of SpeX ($\lambda / \Delta\lambda = 200-2500$). In principle,
use of the routines described below (Sec.\ 4) is limited to data
with $\lambda/\Delta\lambda < 50000$, although data with higher resolving
powers could be corrected if a model with higher resolving power were incorporated
into the code. 
We have used a more general version of the code described below to successfully 
perform telluric corrections on data with resolving power of 5000 and we routinely 
use the code, with an empirical instrument profile $P(\lambda)$, to successfully correct 
low resolution ($\lambda / \Delta\lambda \sim 200$) prism data obtained with SpeX. 
At low resolving powers, the possibility of finding isolated H lines that are 
uncontaminated by atmospheric absorption is greatly diminished and precludes the
construction of the convolution kernel $K(\lambda)$ from the observed data.
At $\lambda / \Delta\lambda \sim 200$, for example, even the Pa $\delta$ line is 
affected by telluric absorption in the wings.

The calculations described above (Sec.\ 2) are easiest to carry out in velocity
coordinates, which is appropriate for a spectrum with constant spectral resolving power.
In addition, the Vega model has a constant resolving power. However, in most 
spectrograph designs, the resolving power varies across the spectral range. 
SpeX is no exception, and the resolving power varies linearly by about 20\% 
across each order (Rayner et al.\ 2002). Fortunately, the spectral resolution 
$\Delta\lambda$ of SpeX is constant to a high degree 
across each order, although the specific value varies from order to order. 
However, in pixel coordinates it is the same value for all orders, given approximately 
by the projected slit width. This implies that, in pixel coordinates, a kernel 
$K(\lambda)$ built from a line in one order can be applied to all other orders. Therefore, 
the code described below carries out the deconvolution in wavelength
coordinates in order to construct the kernel, converts the resulting $K(\lambda)$ 
to pixel coordinates, and then applies this kernel to each order. This saves 
the user from the tedious task of building kernels specific to each order.

Deconvolution procedures are notoriously sensitive to noise.
To minimize the effects of noise on the deconvolution needed to construct 
$K(\lambda)$, we multiply the 
ratio of the Fourier transforms of $O'_{A0V}(\lambda)$ and $I'_{Vega}(\lambda)$
by a tapered window function of the 
form $1/(1 + (\nu/\nu_{max})^{10})$, where $\nu_{max}$ is a reference frequency given
by a multiple of the width of a Gaussian fit to the Fourier transforms. This has the
effect of suppressing high frequency noise. 

%Of course, it is also important to 
%acquire high S/N spectra of the A0V star to begin with.

%Despite this, we have found that no modification
%needs to be made to the convolution kernel in order to subtract cleanly
%the stellar lines across this wavelength range. For spectrographs whose
%resolution, sampling, or dispersion changes more substantially between the
%region around Pa $\delta$ and other wavelengths of interest, the
%convolved spectrum of Vega will need to be modified further to account
%for these variations.

Clearly, as the response spectrum $R(\lambda)$ will be divided into the observed
target spectrum, it is also important to acquire high S/N values in the
observed spectrum of the A0V star in order to achieve accurate and reliable
telluric corrections. In general, we have found that a S/N 
of $\geq 100$ is needed in order to obtain telluric-corrected spectra that 
are limited by the flat-fielding errors and the S/N of the observed target
spectrum. Of course, the accuracy of the final corrected spectra also 
depends on the stability of the atmospheric conditions, the 
proximity (in both time and airmass) of the A0V observations to
those of the target object, and the degree to which the A0V spectrum matches the 
model spectrum of Vega.
However, as we have argued above, small spectral mismatches should have a minor effect
on the generated telluric correction spectrum. 
Fortunately, S/N values of 100 are rather easy to
achieve at resolving powers of $\leq 2500$ on 3-m class telescopes. 
Using SIMBAD we have compiled a list
of A0V stars that are suitable for carrying out the method described 
above. The SpeX Webpage includes a web form that will generate a 
listing of the known A0V stars within a specified angular distance and/or airmass 
difference from any target object. Our experience using SpeX on Mauna Kea indicates
that on particularly dry and stable nights, accurate corrections can be achieved
when the airmass differences and time intervals between the target and standard observations are
as large as 0.1 and 30 to 60 min, respectively. However, for typical conditions, good telluric 
corrections require closer matches in airmass and shorter time separations.

Accurate flux calibrations require knowledge of the near-infrared fluxes or magnitudes
of a set of A0V stars. Unfortunately an accurate and comprehensive set of fluxes and/or
magnitude measurements for A0V stars at these wavelengths is currently not available. Therefore, we have 
chosen to adopt the optical ($B$ and $V$) magnitudes given by SIMBAD and
the reddening curve given by Rieke \& Lebofsky (1985; see also Rieke, Rieke, \& Paul 1989) 
to estimate the flux as a function of wavelength for a given A0V star. In order to
achieve a S/N of about 100 without inordinately long exposure times, 
the telluric A0V star must be reasonably bright; for SpeX observations, A0V stars with $V$
magnitudes of $\sim 5 - 8$ have been found to work best. SIMBAD provides highly accurate
magnitudes for a large number of A0V stars in this brightness range. In addition, the reddening 
of such objects is necessarily small, even in the visible. Hence the precise reddening curve 
adopted has a minor effect on the results. However, we point out that accurate spectroscopic 
flux calibrations at near-infrared wavelengths would clearly benefit from a dedicated 
program of careful $J$, $H$, and $K$ observations of a set of A0V stars. Such a
program could also serve as the basis for establishing a set of near-infrared 
``spectroscopic flux standards'' similar to those available in the optical regime.

Proper flux calibration also requires that the seeing and other slit losses do not change 
substantially between the observations of the target and the A0V star. If the 
apertures used for the spectral extraction are wide, moderate changes in seeing should 
not affect the resulting flux calibrations greatly. 
In practice we find that for typical seeing conditions on Mauna Kea at the IRTF and extraction 
apertures of $\sim 2''$ width, the derived fluxes agree
with those estimated from broad-band magnitudes to within a few percent.

\section{Implementation: xtellcor}
\label{s:implementation}

The method outlined above (Sec.\ 2) has been implemented in a graphical IDL-based
code called {\bf xtellcor}. Here we describe the version of {\bf xtellcor} specifically
designed to operate on SpeX data. A more general version of {\bf xtellcor}
(for non-SpeX data) is similar, but does not construct a convolution kernel $K(\lambda)$ from
the A0V star; rather the user must input values of the parameters describing the
IP, $P(\lambda)$. Both versions of the code are available from the SpeX Website.
Help files are included in the distribution packages to guide the user through the
various steps.

Starting {\bf xtellcor} brings up an IDL widget containing
fields in which the user can enter the information necessary for the
code to run (data directory and file names for the observed A0V and target object spectra; 
see Fig.\ \ref{f:xwidget}). Since near-infrared magnitudes of a large set of
A0V stars are not available, the user must give the $B$ and $V$ mags of the
observed A0V star. These magnitudes are used to redden and scale the model
spectrum of Vega in order to set the flux calibration properly.

At this stage the user can choose to build a
kernel from the observed A0V star, or accept an empirical IP
constructed from the profiles of arc lines in SpeX wavelength calibration images. 
%A cross between implementing a purely
%theoretical and a purely empirical IP, 
The latter is particularly
useful in cases where the IP cannot be constructed from the observed A0V
star (e.g., in the long-wavelength cross-dispersed modes of SpeX which cover the 
1.9-5.5 \mm\ range). 
The arc line profiles have been fitted with an empirical function of the form
\begin{equation}
 P(x) = C \cdot [{\rm erf}((x+a)/b)  - {\rm erf}((x-a)/b)],
\end{equation}
where $C$ is a normalization constant and $a$ and $b$ determine the
width of the profile in pixel ($x$) coordinates.
This functional form provides an excellent fit to the observed profiles
of the arc lines over all the orders of SpeX and for slit widths of
0.3$''$, 0.5$''$, 0.8$''$, and 1.6$''$.
By adopting a fit, rather than the observed line profiles themselves,
errors in the IP resulting from undersampling, low S/N, etc., are minimized.
The best fit values of $a$ and $b$ for each slit width are stored in a
separate data file. 

If the user chooses to build the kernel from the A0V star, a second window is
generated containing the observed spectrum in the requested order. 
The Pa $\delta$ line is located in order 6 of the 0.8-2.4 \mm\ cross-dispersed
mode of SpeX. The user can
then interactively select continuum regions to be fit with a polynomial
of a desired degree. Once an adequate fit to the continuum in the
vicinity of the spectral line is obtained, the normalized spectrum (observed
divided by the continuum fit) is displayed (Fig.\ \ref{f:xkern}).
The user then selects the region
of the normalized spectrum to use in the deconvolution and
the code proceeds to scale both the model line spectrum and the normalized model
continuum spectrum by the ratio of the observed
equivalent width to the model equivalent width of the selected line (Fig.\ \ref{f:xkern}). 
The relative velocity shift between the observed
spectrum and the model is determined and the kernel is generated via deconvolution
after the model has been shifted. Our experience has shown that with careful choice 
of the normalization and deconvolution
regions, a kernel can be routinely constructed that yields maximum deviations of the
convolved model from the observed line profile of less than 2\% over the selected
wavelength range. We typically achieve maximum deviations of less than 1.5\% 
and RMS deviations of less than 0.75\% across the Pa $\delta$ line for data with
$\lambda/\Delta\lambda = 2000$ and S/N $\geq 100$.

The observed spectrum of the A0V star is divided by 
the line model after the latter has been convolved with the kernel (either the semi-empirical
IP, or one built from the observed A0V star) and the result is displayed for each order
(Figs.\ \ref{f:xscale} and \ref{f:xscale2}). This is the first version of the 
response spectrum. At this stage the user can adjust the velocity shift of the
model and modify the strengths of the various H lines to better match those of
the observed A0V star. Mismatches in line strengths and velocities are easily
seen in the telluric correction spectrum as ``absorption'' or ``emission'' features
at the location of the H lines. The user can adjust the individual H line strengths
of the model by varying the scale factors applied to the model line equivalent widths 
until residual features are minimized in the response spectrum in the regions around the 
lines (see Figs.\ \ref{f:xscale} and \ref{f:xscale2}). This can be done
graphically using the mouse until residual features
are minimized in the regions around the lines. A representative spectrum of
the atmospheric transmission is displayed to guide the user in this task. (As an 
additional aid to the user,
this atmospheric transmission curve can be multiplied by the estimated
throughput curve of SpeX.) 
The code also includes a procedure to determine the scale factor for a single
isolated line automatically. This procedure requires two points on either side of
a line, fits a Gaussian plus a continuum to the response spectrum and adjusts the strength
of the model line until the height of the Gaussian is minimized. 
Most of the H lines can be easily removed from the response/telluric correction spectrum at
this stage. However, we have found that the Pa 12, Pa 13 and Pa 14 lines (and occasionally
the Pa 11 and Pa 15 lines) almost always leave residual features in the telluric correction 
spectrum that cannot be eliminated, no matter how the line strengths
are scaled. This indicates that the Vega model we are using does not provide a good
match to the profiles of these lines in the spectra of real A0V stars. Once an
improved Vega model is available, we will incorporate it into the code. Fortunately,
even in the vicinity of these discrepant lines, the residual featuress are generally less than
a few percent and therefore do not result in large errors in the final spectrum.  

The resulting set of line strength scale factors are joined with a spline fit to
generate a line scaling spectrum which is multiplied by the line model. The normalized
continuum (divided by the spline fit) is also multiplied by this line scaling curve.
This has the effect of modifying the jumps seen in the continuum model.
Since the highest order lines rarely require modification of their strengths,
the jumps are normally scaled by the equivalent width ratio determined
from the line used to build the kernel. The scaled line model 
is then used to regenerate the response spectrum. When the user is satisfied
with the modifications to the model, the final telluric correction spectrum is generated.
%Before multiplying the target object spectrum by the telluric absorption spectrum, the latter 
This can then be shifted in pixels in order to align the telluric absorption features
with those seen in the target object spectrum. This procedure should minimize any wavelength
shifts due to flexure between the observations of the target object and
the standard star. The code automatically computes
the best pixel shift using a region of the spectra selected by the user (Fig.\ \ref{f:xshift}).
The shifts are usually less than $0.5$ pixels for standard stars observed within 
10 degrees of the target object. The shift value is then applied to all the orders of the telluric
correction spectrum.
Finally, the observed target object spectrum is multiplied by the shifted telluric correction spectrum, 
order by order, and the final corrected object spectrum is generated 
(Fig.\ \ref{f:xoutput}). The user has a choice of units for the final output spectrum. 
The user can also choose to save both the 
telluric correction spectrum itself, as well as the modified Vega spectrum. The 
former is useful if a single A0V star has been observed as a telluric standard 
for multiple target objects; a separate routine allows this previously constructed
telluric correction spectrum to be read in and used to correct the observed
spectra of other objects. 

\section{Application and Examples}
\label{s:application}

The response spectra
generated by {\bf xtellcor} for the various short-wavelength cross-dispersed orders of SpeX
are shown in Fig.\ \ref{f:xresp}(a-f) and Fig.\ 1 of Cushing et al.\ (2002a).
As can be seen from these figures, the features seen in the response spectra match extremely
well with those seen in the theoretical atmospheric spectrum, and the stellar H lines seen in the
observed A0V spectra are removed to a very high accuracy.
Small residuals can be seen in Fig.\ \ref{f:xresp}(e) at the locations of the 
Pa 11 - Pa 15 lines.
Aside from these features, however, the accuracy of the telluric corrections is limited only by
the S/N of the input A0V spectrum. The latter is determined by the exposure time, flat-fielding
errors, instrument throughput, and the atmospheric transmission itself. (Clearly, the derived 
telluric correction spectrum will be highly uncertain in those wavelength regions where the 
atmosphere is so opaque that the observed spectra have S/N close to zero.) 
Of course, as explained above, the accuracy of the final corrected target spectra also depends on 
the differences in time and airmass between the observations of the A0V star and the target object,
as well as the S/N of the target spectrum.

Examples of typical output spectra generated by {\bf xtellcor} can be seen in Fig.\ 
\ref{f:xoutput} (for HD 223385, A3Ia), Fig.\ \ref{f:xfinalo} (for HD 206267, O6.5V((f)),
and Fig.\ \ref{f:xfinalg} (for HD 16139, G8IIa), as well as in the papers by Cushing et al.\ (2002a) 
and Rayner et al.\ (2002). The differences in airmass and time of observations between the 
objects shown in these figures and their telluric A0V standards were $\leq 0.05$ and $\leq 20$ min,
respectively, in all cases. However, the atmospheric conditions (humidity, seeing, transparency) were variable
during the observations. The S/N values of the A0V spectra were always $\geq 100$. 
As the figures demonstrate, the method enables us to recover the target spectrum even in regions
of relatively low atmospheric transmission. 
We note that the equivalent widths of the intrinsic stellar
absorption and emission features seen in the spectra in Figs.\ \ref{f:xoutput} and \ref{f:xfinalo} 
agree well with those measured for these spectral classes by Hanson et al.\ (1996)
in the $K$ band. 

\section{Summary}
\label{s:summary}

We have described a technique for correcting observed near-infrared
medium-resolution spectra for telluric absorption. Designed specifically
for data obtained with SpeX at the IRTF, the method makes use
of a high S/N ($\geq 100$) spectrum of an A0V star observed near in time and 
close in airmass to the object of interest, and uses a theoretical model of
Vega to remove the intrinsic H absorption features seen in the stellar
spectrum. The method can regularly generate telluric correction spectra accurate
to better than 2\% in the vicinity of the intrinsic stellar H lines and
significantly better than that at other wavelengths.

We have demonstrated the application of this method on spectra
obtained with SpeX. A set of graphical, easy-to-use,
IDL-based routines (part of a general SpeX spectral reduction package
called Spextool\_extension v.\ 1.1), implements the procedures described in this paper
and can be found on the SpeX webpage at the IRTF website \\
({\tt http://irtfweb.ifa.hawaii.edu/Facility/spex/}). A general version of the code, useful
for performing telluric corrections on spectra obtained with other near-infrared
spectrographs with resolving powers of $\lambda/\Delta\lambda < 50000$ is also available. In addition, 
a form that generates a list of the A0V stars listed in
the SIMBAD database near a set of specified coordinates is also posted on the website.

%%%%%%%%%%%%%%%
% Acknowledgments
%%%%%%%%%%%%%%%

\acknowledgments

We thank R.\ Kurucz for providing the high resolution model of
Vega and J.\ Tonry for suggestions regarding the form of the
instrument profiles. We also thank the SpeX team for their expert 
construction of an invaluable instrument and the staff at the IRTF for
their assistance at the telescope. This research has made use of the
SIMBAD database, operated at CDS, Strasbourg, France.

%%%%%%%%%%%%%%%
% Appendices
%%%%%%%%%%%%%%%

% clearpage
% Here: Appendix text, if any

\clearpage

%%%%%%%%%%%%%%%
% Use a small baselineskip for the references, unless in submission mode.
%%%%%%%%%%%%%%%

\ifsubmode\else
\baselineskip=10pt
\fi

%%%%%%%%%%%%%%%
% Reference List
%%%%%%%%%%%%%%%

\clearpage

%%%%%%%%%%%%%%%
% Change back to the regular baselineskip, if necessary
%%%%%%%%%%%%%%%

\ifsubmode\else
\baselineskip=14pt
\fi

%%%%%%%%%%%%%%%
% Figure Captions
%%%%%%%%%%%%%%%

\newcommand{\figcapabsorp}{
The normalized model spectrum of Vega, smoothed to a resolving power
of 2000 (thick line, left hand axis).
The theoretical atmospheric transmission around the Pa $\delta$ line 
(1.00494 \mm) at airmass 1.2 for typical conditions on Mauna Kea, smoothed
to a resolving power of 2000 (thin line, right hand axis). 
Note that the Pa $\delta$ line
is located in a region where atmospheric absorption is negligible.
\label{f:absorption}}

\newcommand{\figcapxwidget}{Control panel widget for {\bf xtellcor}.
\label{f:xwidget}}

\newcommand{\figcapxkern}{Determination of the convolution kernel from the 
Pa $\delta$ line in the normalized spectrum of an A0V star. The vertical 
dashed lines denote the region of the spectrum used in the deconvolution to
generate the kernel. Note that the
residuals (seen in the bottom panel) have an RMS deviation of much less than
1 \% .
\label{f:xkern}}

\newcommand{\figcapxscale}{Determination of the H line scale factors for the Vega model
in order 4 of SpeX. The upper panel shows the scale factors for each of the H 
absorption lines, while the bottom shows the response spectrum. The yellow
curve is the theoretical atmospheric transmission for an airmass of 1.2 and standard
conditions on Mauna Kea. 
\label{f:xscale}}

\newcommand{\figcapxscaleb}{Same as Fig.\ 4 for order 6 of SpeX. Note the
small residual features in the response spectrum at the locations of Pa 12 and Pa 13.
\label{f:xscale2}}

\newcommand{\figcapxshift}{Determination of the shift between the telluric correction
spectrum and the observed target object spectrum. The vertical dashed lines denote the 
region used to determine the pixel shift. The solid white line is the observed object
spectrum, while the green line is the system response (the inverse of the telluric 
correction spectrum). The lower panel shows the telluric-corrected spectrum for this order.
\label{f:xshift}}

\newcommand{\figcapxoutput}{The final telluric-corrected spectrum of HD 223385 (A3Ia).
\label{f:xoutput}}

\newcommand{\figcapxresp}{Response curves generated by {\bf xtellcor} for the various
short-wavelength cross-dispersed orders of SpeX (8{\it a}: order 3; 8{\it b}: order 4; 
8{\it c}: order 5; 8{\it d}: order 6; 8{\it e}: order 7; 8{\it f}: order 8). In each plot, 
the curves shown (from
bottom to top) are (i) the observed spectrum of an A0V star (HD 223386, $V=6.328$),
the response curve generated with {\bf xtellcor}, and a theoretical atmospheric
transmission spectrum computed for typical conditions on Mauna Kea, an airmass similar
to that of the observations, and a resolving power of 2000. The intrinsic stellar H
lines of the Paschen and Brackett series are identified. Small residuals are seen
in the response curves at the locations of the Pa 11-15 lines. Note that the theoretical
atmospheric transmission spectrum does not extend below 0.85 \mm.
\label{f:xresp}}

\newcommand{\figcapxfinalo}{Final telluric-corrected spectra of an O6.5V((f)) star (HD 206267),
generated by {\bf xtellcor} for the various short-wavelength cross-dispersed orders of SpeX 
(9{\it a}: order 3; 9{\it b}: order 4; 9{\it c}: order 5; 9{\it d}: order 6; 9{\it e}: order 7).
In each plot, the spectra are ordered from bottom to top as follows: the observed stellar O6.5V((f))
spectrum, the response curve constructed by {\bf xtellcor} from the A0V telluric standard,
the telluric-corrected O6.5V((f)) spectrum, and the representative theoretical atmospheric transmission
spectrum (for reference).
\label{f:xfinalo}}

\newcommand{\figcapxfinalg}{Final telluric-corrected spectra of a G8IIIa~ star (HD 16139),
generated by {\bf xtellcor} for the various short-wavelength cross-dispersed orders of SpeX 
(10{\it a}: order 3; 10{\it b}: order 4; 10{\it c}: order 5; 10{\it d}: order 6; 10{\it e}: order 7).
In each plot, the spectra are ordered from bottom to top as follows: the observed stellar G8IIa
spectrum, the response curve constructed by {\bf xtellcor} from the A0V telluric standard,
the telluric-corrected G8IIIa spectrum, and the representative theoretical atmospheric transmission
spectrum (for reference).
\label{f:xfinalg}}

%%%%%%%%%%%%%%%
% Figures (in submission mode captions only, unless \printfigtrue)
%%%%%%%%%%%%%%%

\ifsubmode
\figcaption{\figcapabsorp}
%\figcaption{\figcapabsorp}
\figcaption{\figcapxwidget}
%\figcaption{\figcapxnorm}
\figcaption{\figcapxkern}
\figcaption{\figcapxscale}
\figcaption{\figcapxscaleb}
\figcaption{\figcapxshift}
\figcaption{\figcapxoutput}
\figcaption{\figcapxresp}
\figcaption{\figcapxfinalo}
\figcaption{\figcapxfinalg}
\clearpage
\else\printfigtrue\fi

%%%%%%%%%%%%%%%
% Tables
%%%%%%%%%%%%%%%
 
\clearpage
\ifsubmode\pagestyle{empty}\fi
 
%%% TABLE 1 %%%

\begin{deluxetable}{rrr}
\tablenum{1}
\tabletypesize{\small}
\tablewidth{0pt}
\tablecaption{Cumulative Number of A0V and G2V Stars listed in the SIMBAD Database, as of July 2000,
as a function of $V$ band magnitude.}
\tablehead{
\colhead{V (mag)} &
\colhead{$N({\rm A0V})~$} &
\colhead{$N({\rm G2V})~$}
}\startdata
$< ~7$~ &  364~ &   50~ \\
$< ~8$~ &  740~ &  171~ \\
$< ~9$~ & 1687~ &  459~ \\
$< 10$~ & 3633~ &  993~ \\
$< 11$~ & 4557~ & 1165~ \\
\enddata
\label{tab:numbers}
\end{deluxetable}
\clearpage

\ifprintfig

%%% FIGURES %%%

\clearpage
\begin{figure}
%\epsfxsize=12.0truecm
%\plotone{vega_atmos.ps}
\plotone{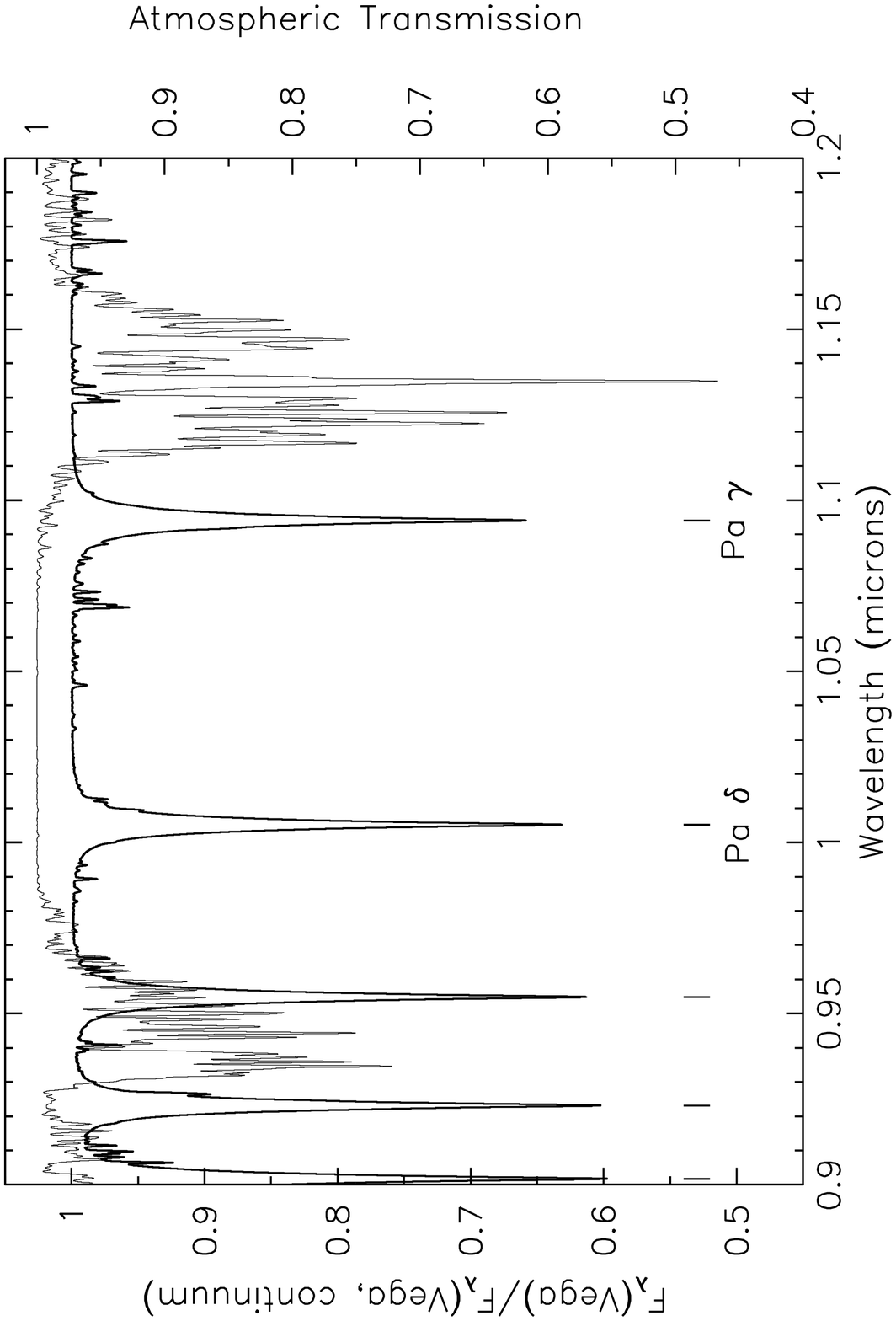}
\ifsubmode
\vskip3.0truecm
\setcounter{figure}{0}
\addtocounter{figure}{1}
\centerline{Figure~\thefigure}
\else\vskip-0.3truecm\figcaption{\figcapabsorp}\fi
\end{figure}

\clearpage
\begin{figure}
%\epsfxsize=12.0truecm
%\plotone{xtellcor_widget.ps}
\plotone{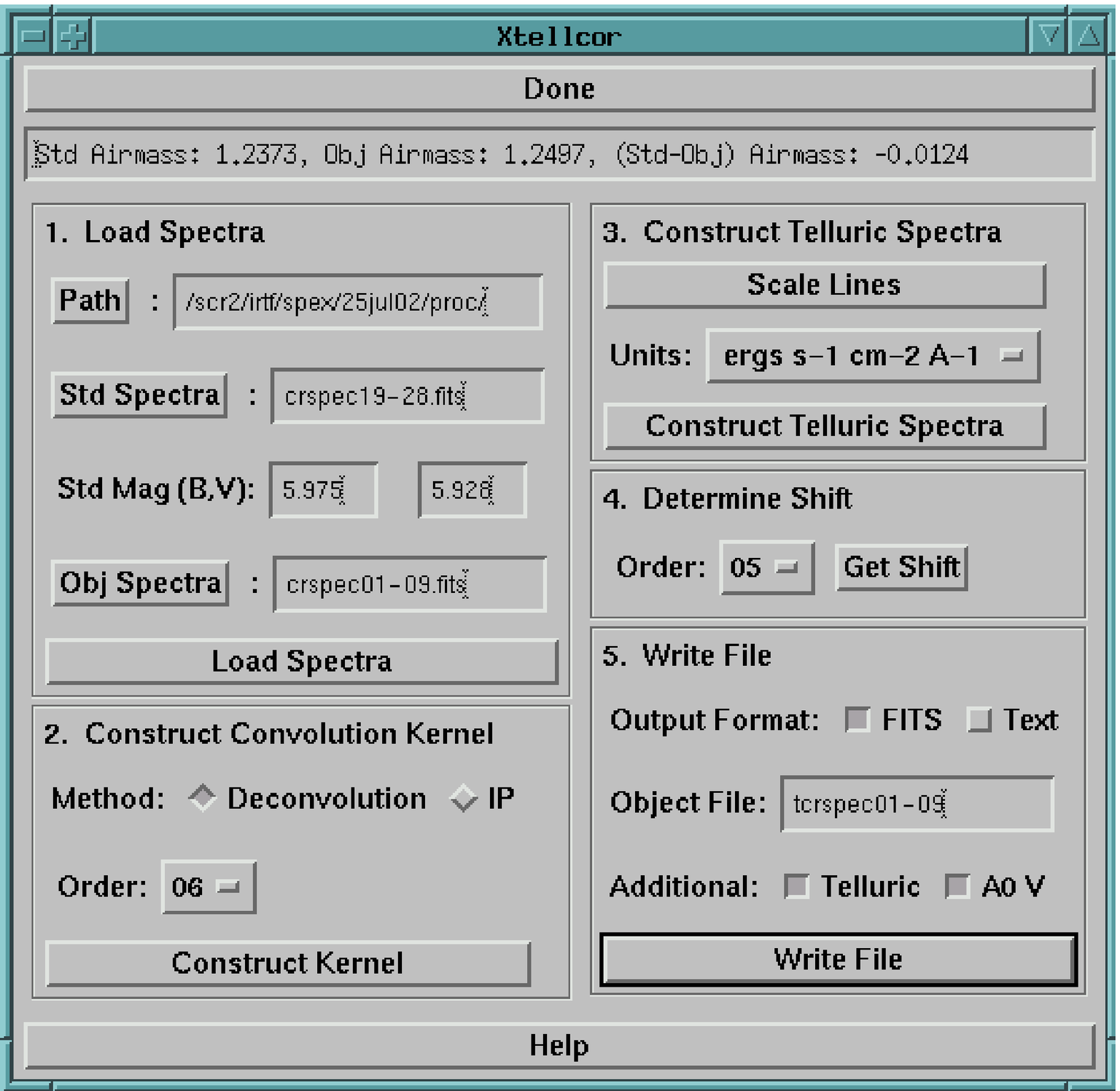}
\ifsubmode
\vskip3.0truecm
\addtocounter{figure}{1}
\centerline{Figure~\thefigure}
\else\vskip0.3truecm\figcaption{\figcapxwidget}\fi
\end{figure}

\clearpage
\begin{figure}
%\epsfxsize=12.0truecm
%\plotone{xtellcor_kernel2.ps}
\plotone{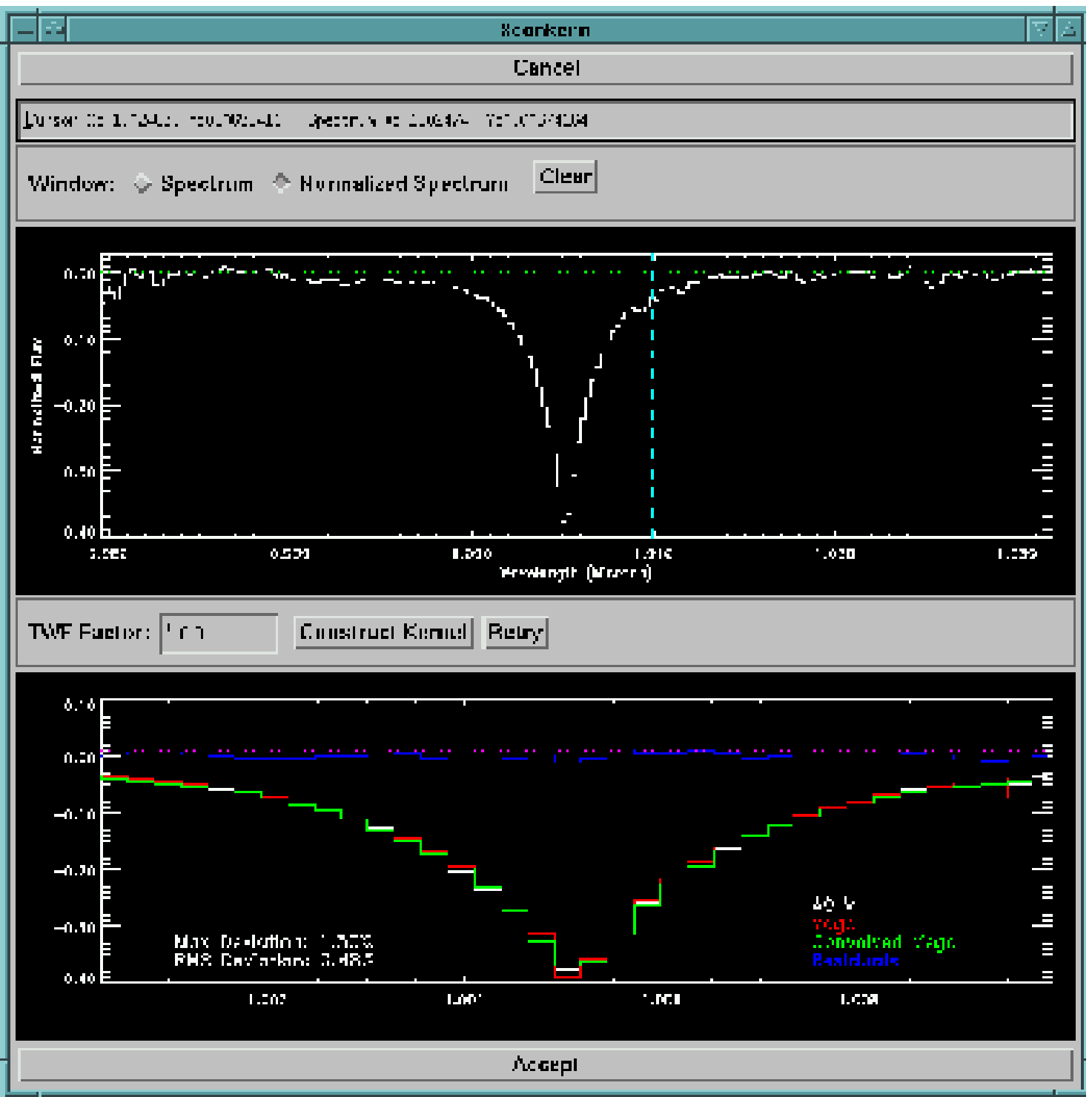}
\ifsubmode
\vskip3.0truecm
\addtocounter{figure}{1}
\centerline{Figure~\thefigure}
\else\vskip0.3truecm\figcaption{\figcapxkern}\fi
\end{figure}

\clearpage
\begin{figure}
%\epsfxsize=12.0truecm
%\plotone{xtellcor_scale.ps}
\plotone{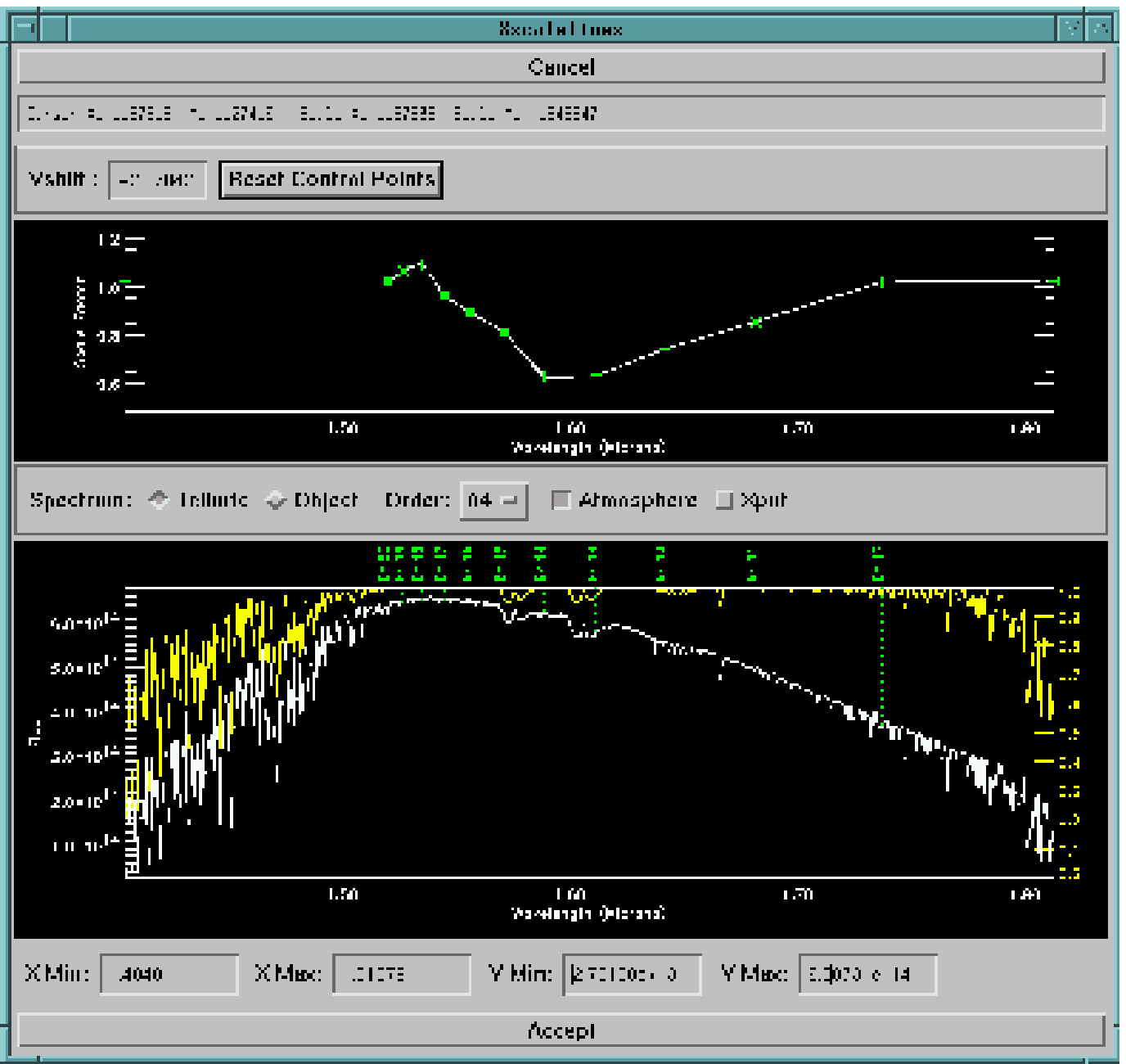}
\ifsubmode
\vskip3.0truecm
\addtocounter{figure}{1}
\centerline{Figure~\thefigure}
\else\vskip0.3truecm\figcaption{\figcapxscale}\fi
\end{figure}

\clearpage
\begin{figure}
%\epsfxsize=12.0truecm
%\plotone{xtellcor_scale2.ps}
\plotone{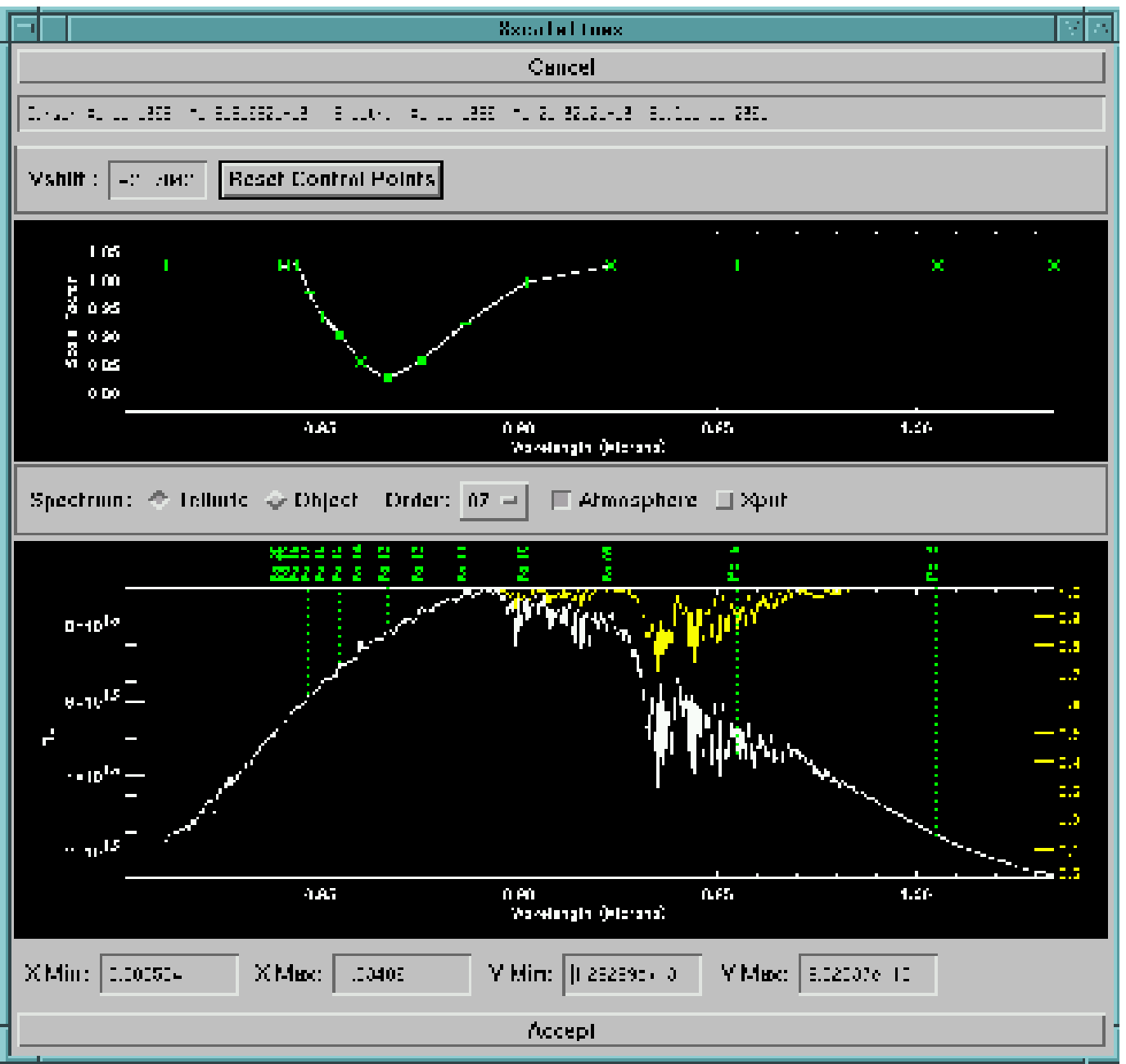}
\ifsubmode
\vskip3.0truecm
\addtocounter{figure}{1}
\centerline{Figure~\thefigure}
\else\vskip0.3truecm\figcaption{\figcapxscaleb}\fi
\end{figure}

\clearpage
\begin{figure}
%\epsfxsize=12.0truecm
%\plotone{xtellcor_shift.ps}
\plotone{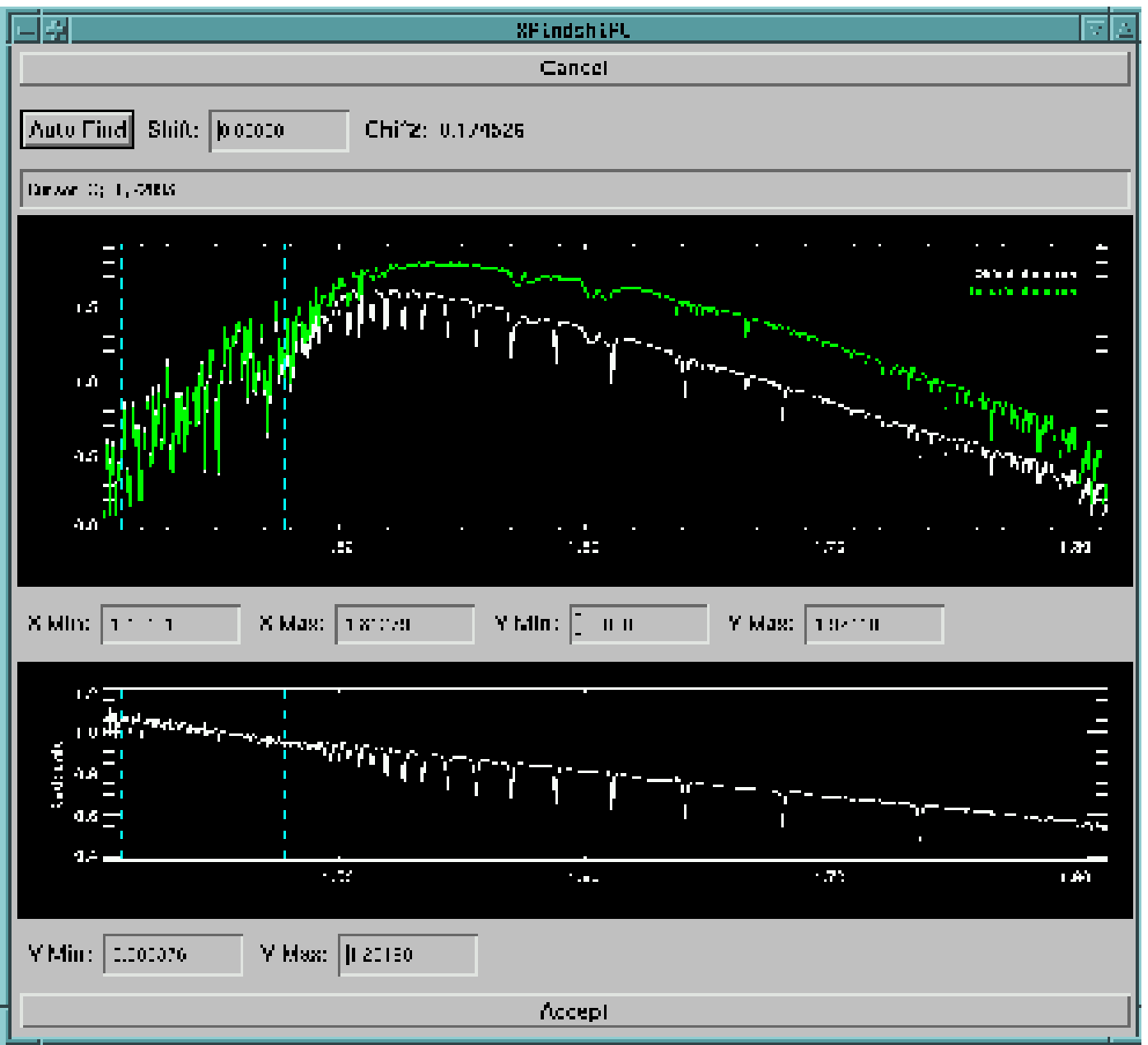}
\ifsubmode
\vskip3.0truecm
\addtocounter{figure}{1}
\centerline{Figure~\thefigure}
\else\vskip0.3truecm\figcaption{\figcapxshift}\fi
\end{figure}

\clearpage
\begin{figure}
%\epsfxsize=12.0truecm
%\plotone{xtellcor_output.ps}
\plotone{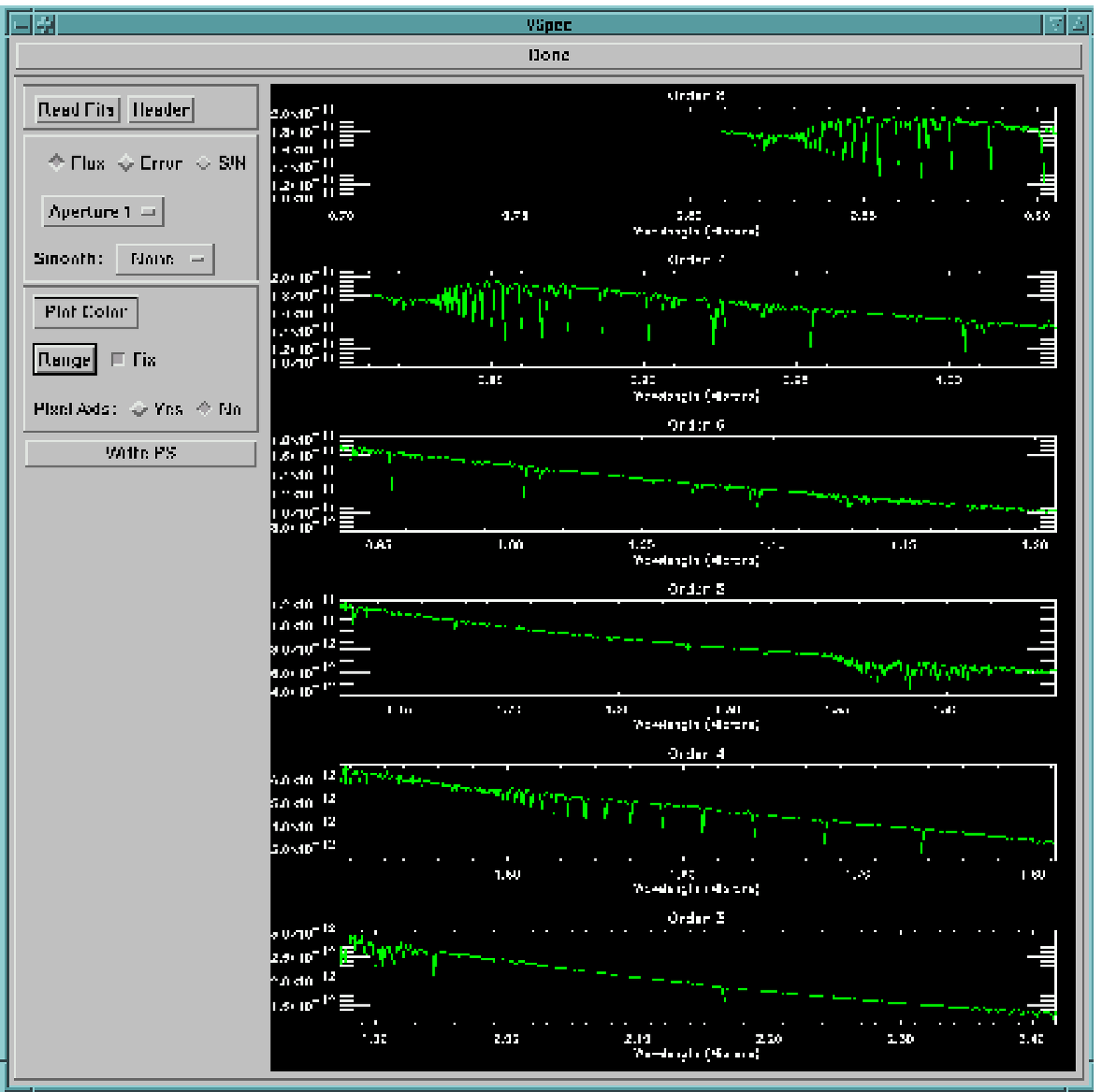}
\ifsubmode
\vskip3.0truecm
\addtocounter{figure}{1}
\centerline{Figure~\thefigure}
\else\vskip0.3truecm\figcaption{\figcapxoutput}\fi
\end{figure}

\clearpage
\begin{figure}
%\epsfxsize=12.0truecm
%\plotone{xtellcor_resp_order3_2.ps}
\vskip-1.5truecm
\plotone{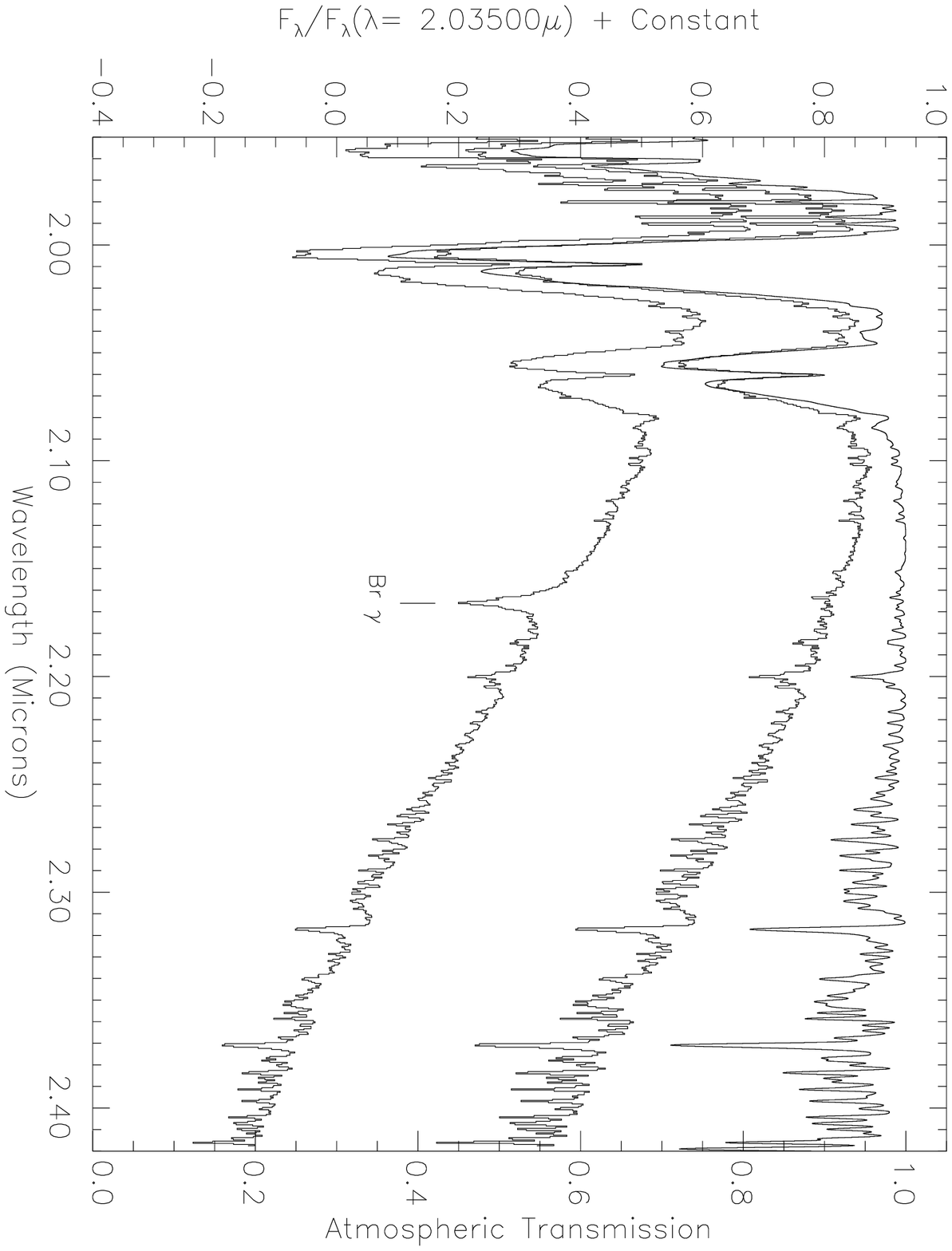}
\ifsubmode
\vskip0.5truecm
\addtocounter{figure}{1}
\centerline{Figure~\thefigure {\it a}}
\else\vskip-2.3truecm\figcaption{\figcapxresp}\fi
\end{figure}

\clearpage
\begin{figure}
%\epsfxsize=12.0truecm
%\plotone{xtellcor_resp_order4.ps}
\plotone{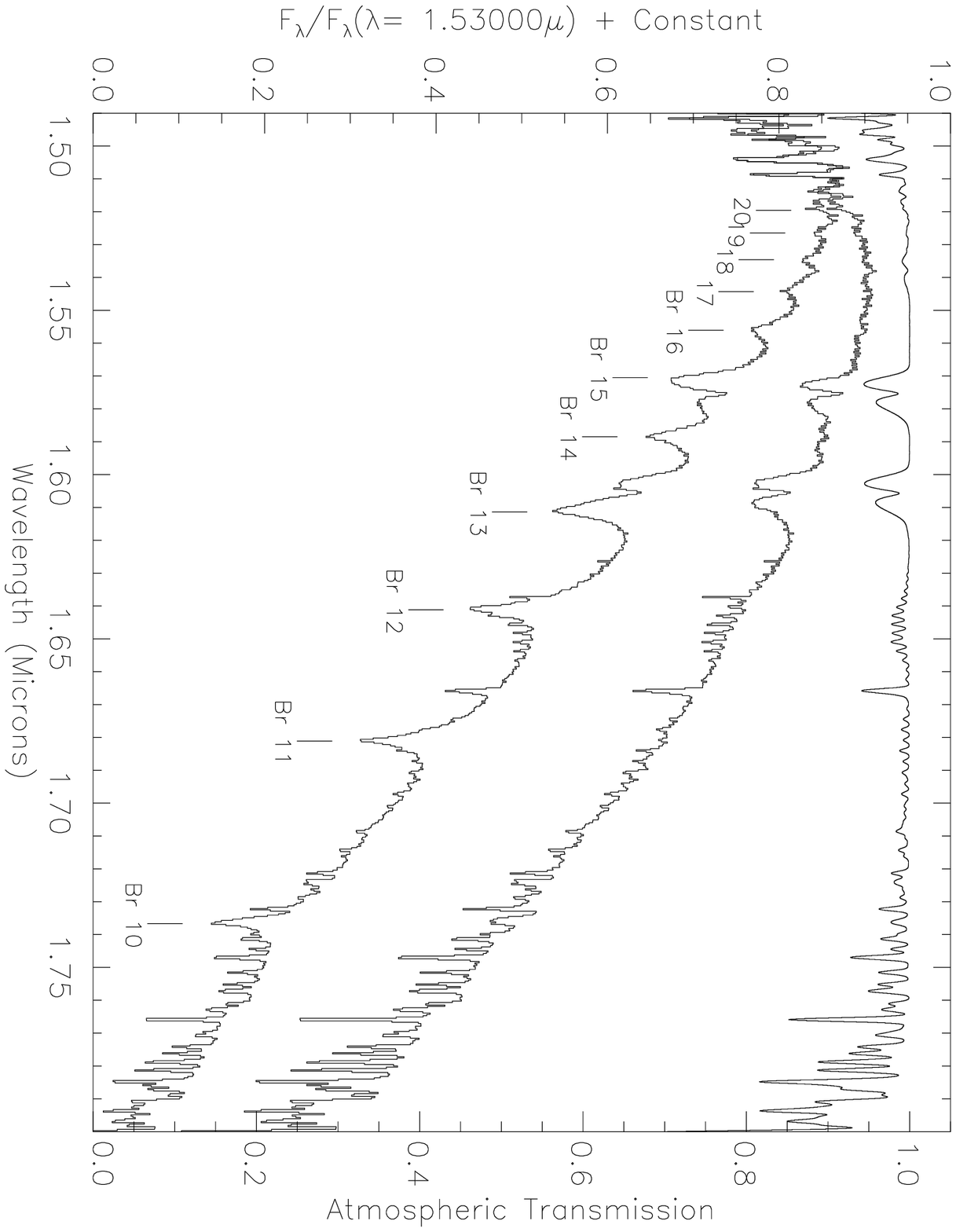}
\addtocounter{figure}{0}
\ifsubmode
\vskip0.5truecm
\else\vskip-0.3truecm\fi
\centerline{Figure~\thefigure {\it b}}
\end{figure}

\clearpage
\begin{figure}
%\epsfxsize=12.0truecm
%\plotone{xtellcor_resp_order5.ps}
\plotone{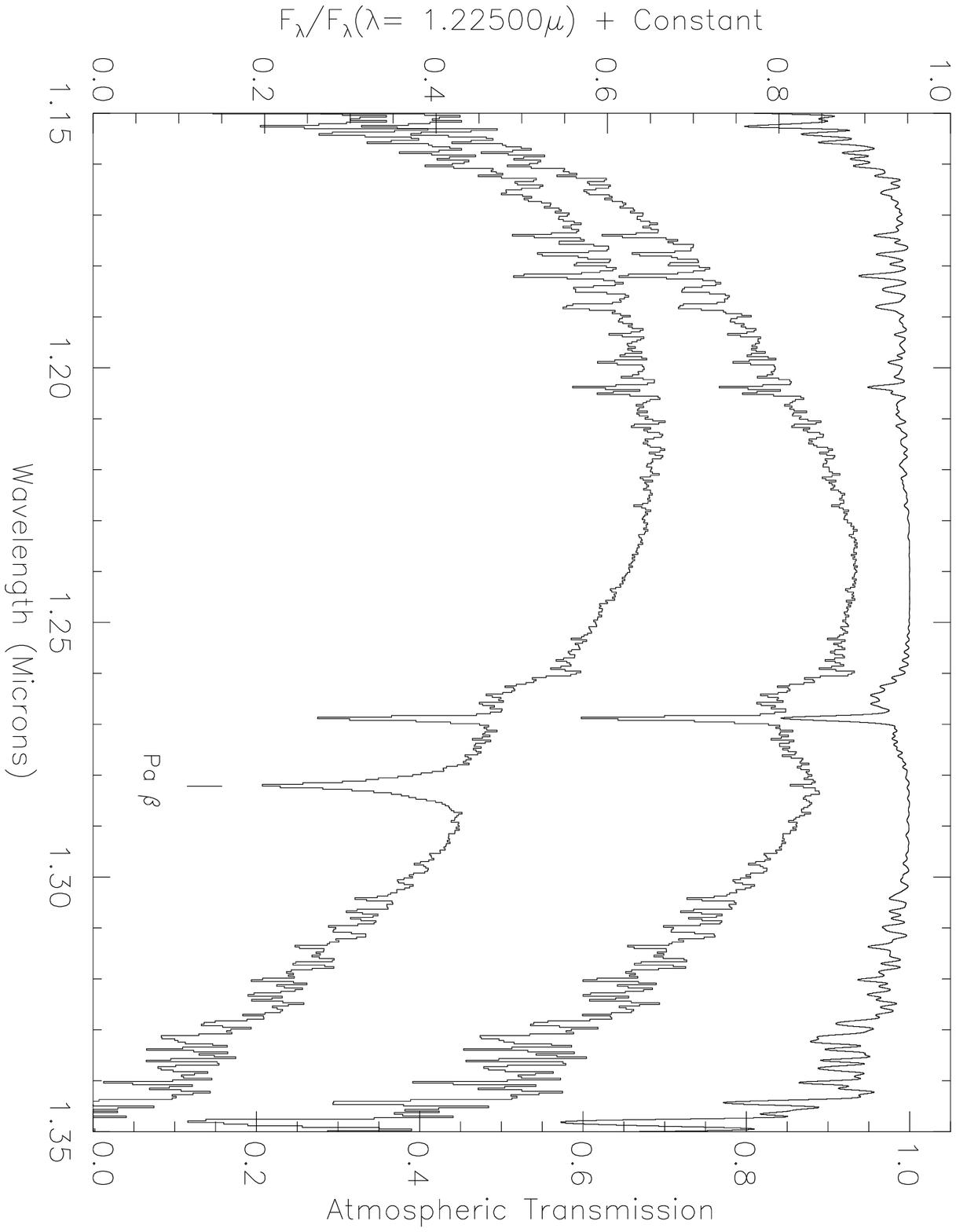}
\addtocounter{figure}{0}
\ifsubmode
\vskip0.5truecm
\else\vskip-0.3truecm\fi
\centerline{Figure~\thefigure {\it c}}
\end{figure}

\clearpage
\begin{figure}
%\epsfxsize=12.0truecm
%\plotone{xtellcor_resp_order6.ps}
\plotone{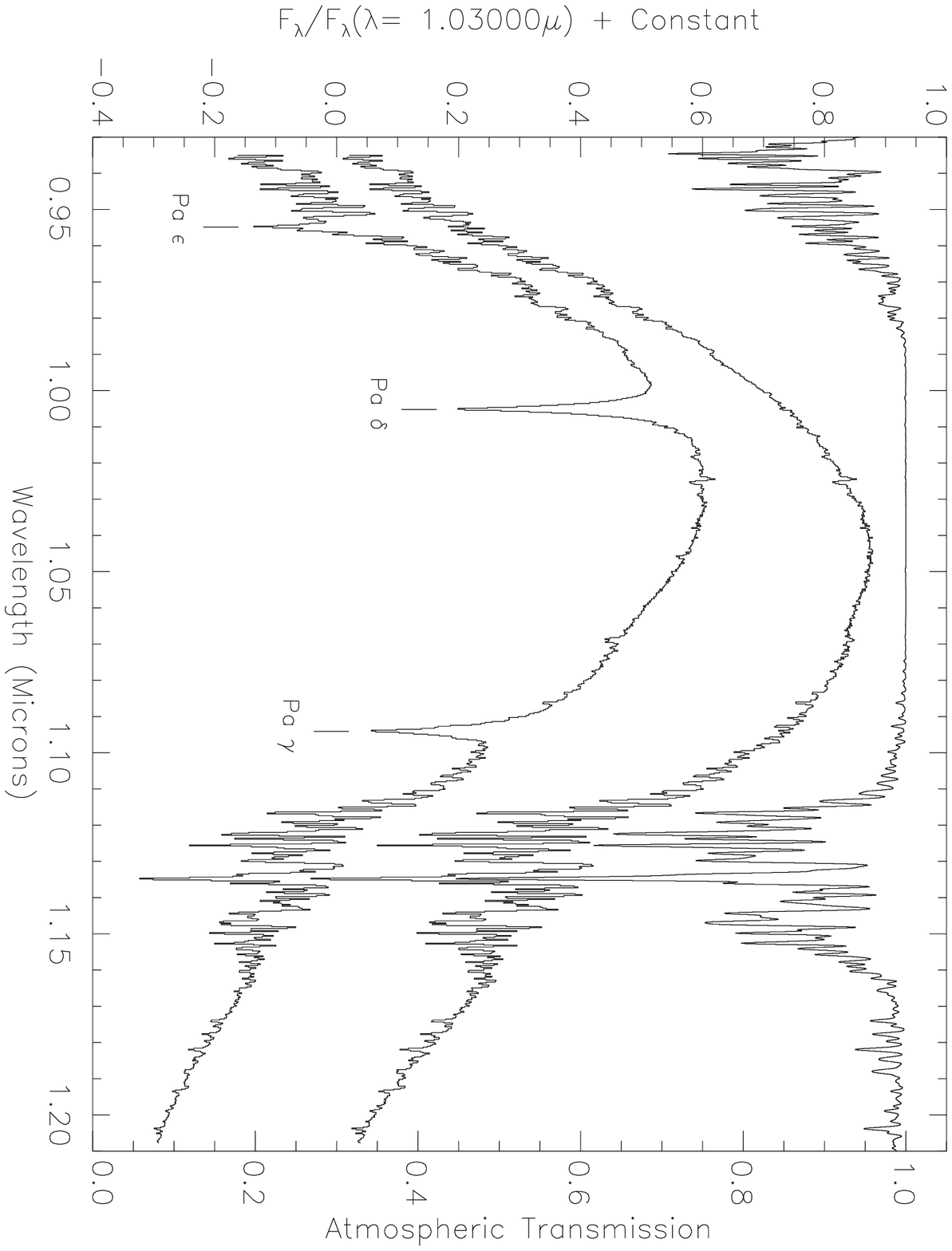}
\addtocounter{figure}{0}
\ifsubmode
\vskip0.5truecm
\else\vskip-0.3truecm\fi
\centerline{Figure~\thefigure {\it d}}
\end{figure}

\clearpage
\begin{figure}
%\epsfxsize=12.0truecm
%\plotone{xtellcor_resp_order7.ps}
\plotone{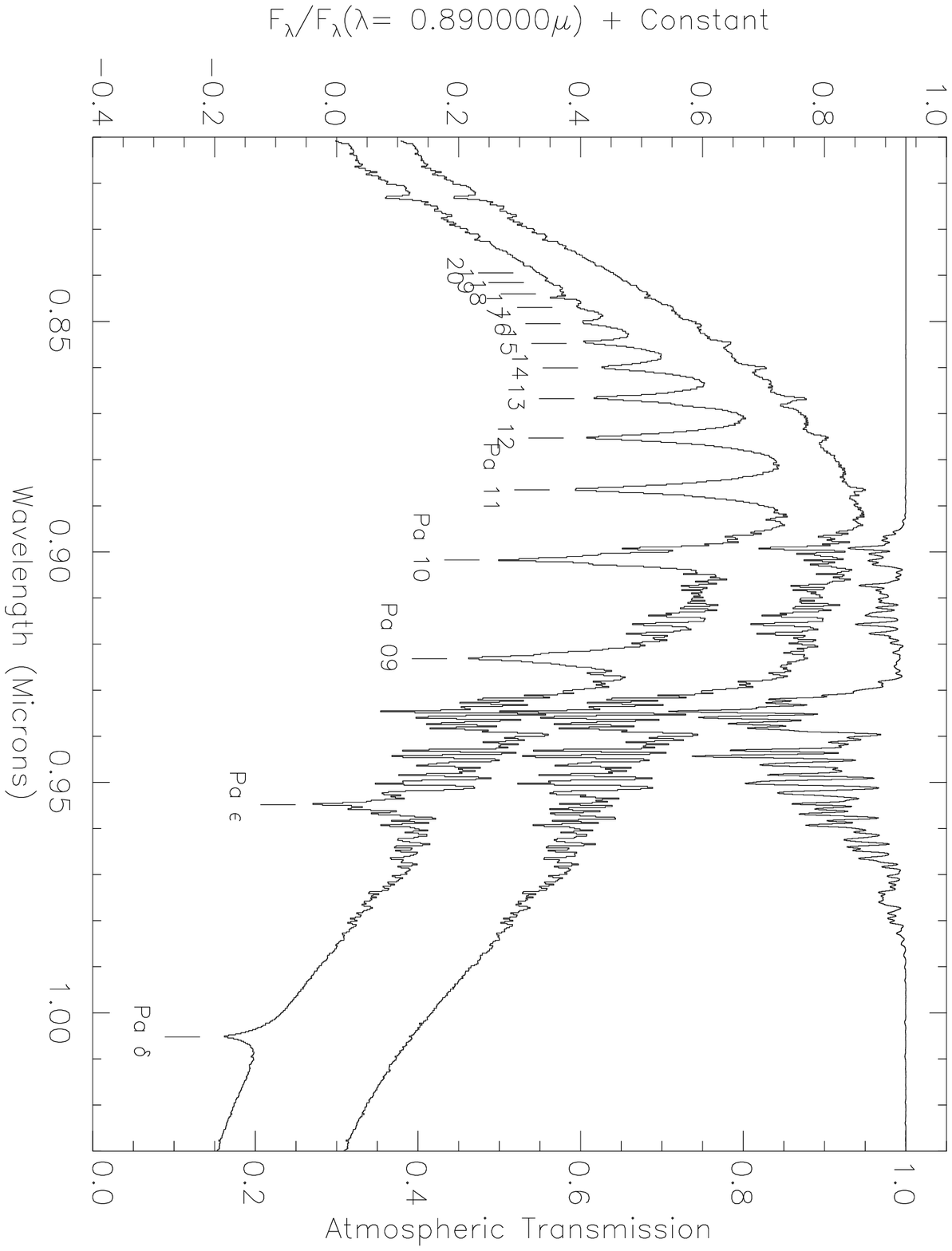}
\addtocounter{figure}{0}
\ifsubmode
\vskip0.5truecm
\else\vskip-0.3truecm\fi
\centerline{Figure~\thefigure {\it e}}
\end{figure}

\clearpage
\begin{figure}
%\epsfxsize=12.0truecm
%\plotone{xtellcor_resp_order8.ps}
\plotone{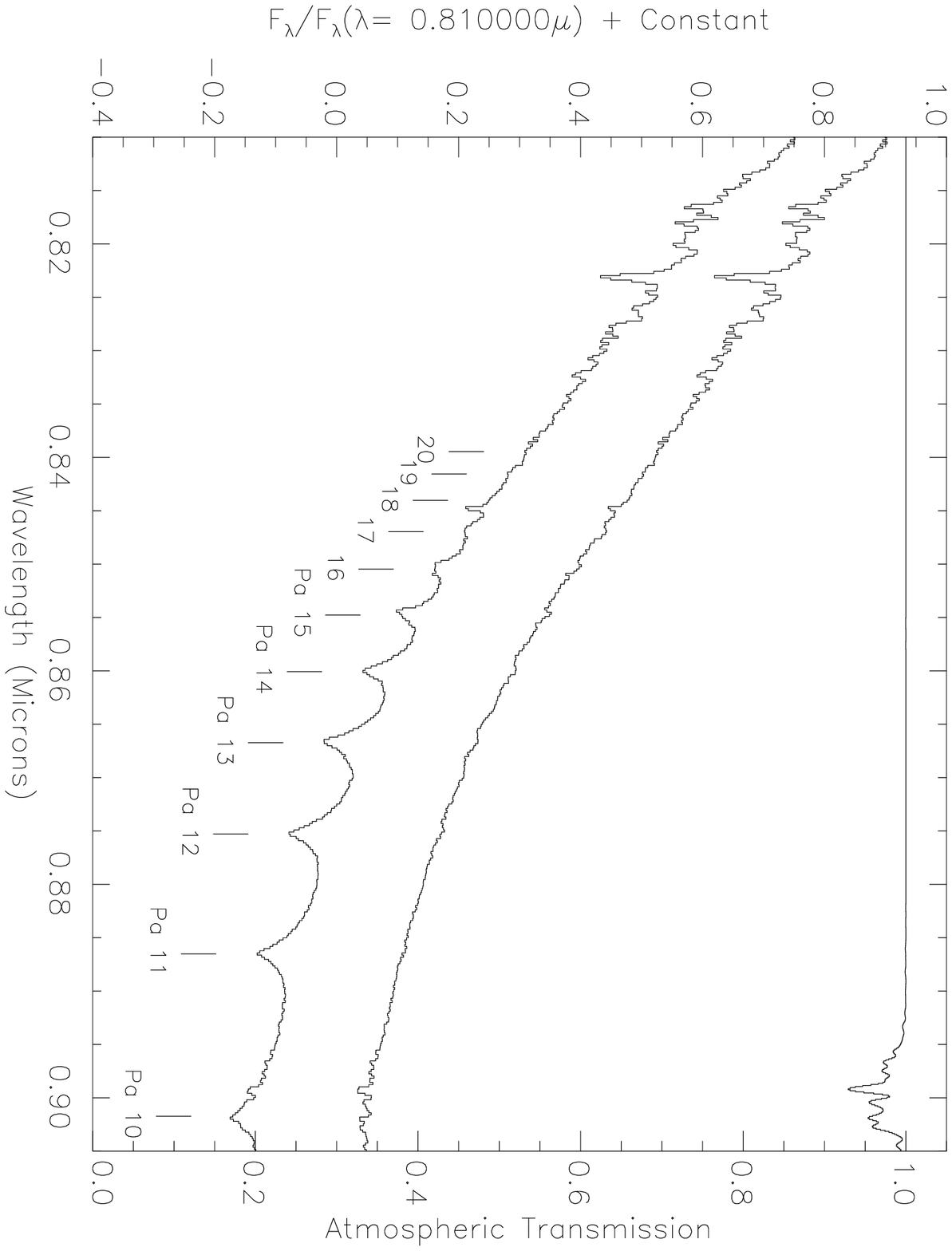}
\addtocounter{figure}{0}
\ifsubmode
\vskip0.5truecm
\else\vskip-0.3truecm\fi
\centerline{Figure~\thefigure {\it f}}
\end{figure}

\clearpage
\begin{figure}
%\epsfxsize=12.0truecm
%\plotone{xtellcor_final_order3_O6.5v.ps}
\vskip-1.5truecm
\plotone{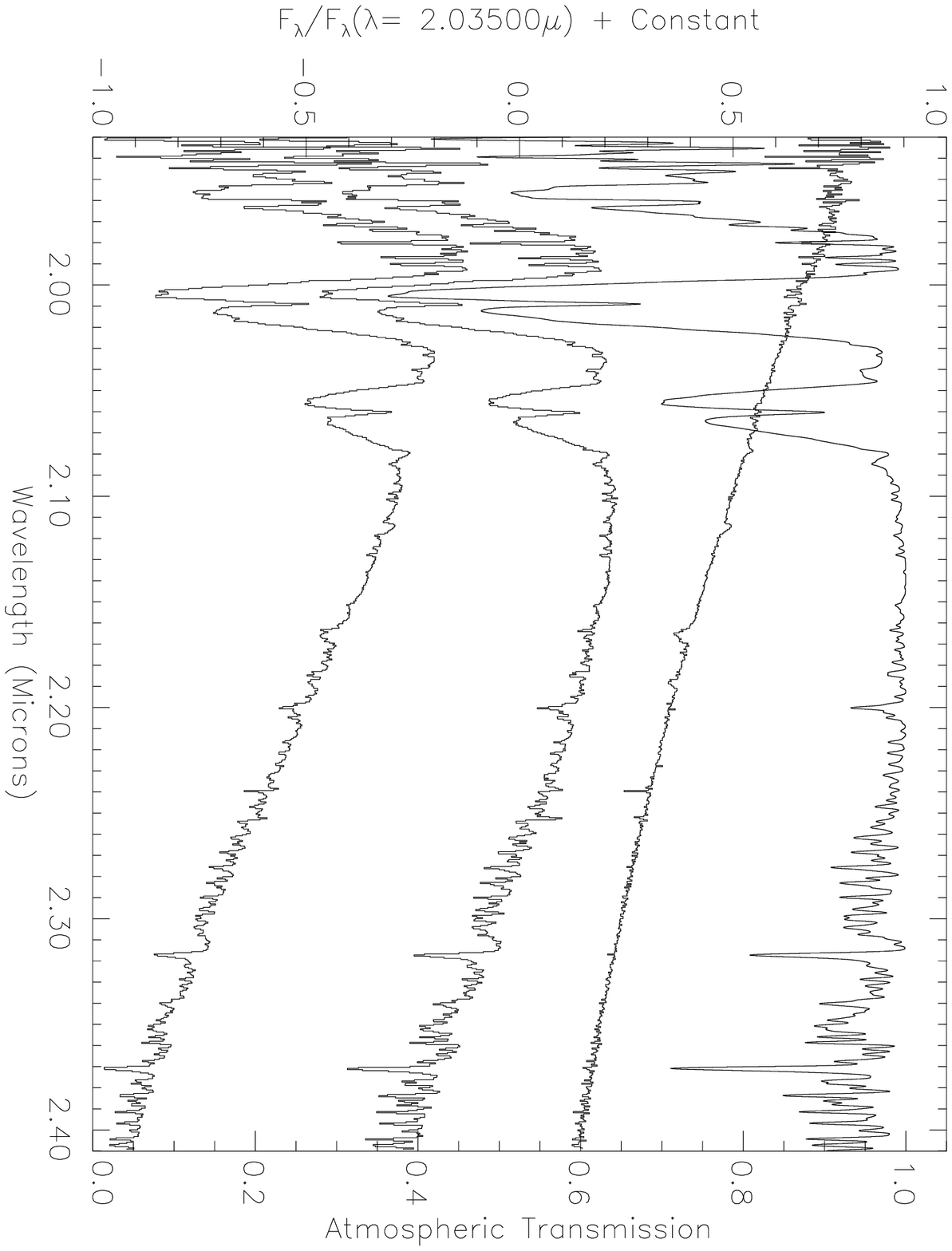}
\ifsubmode
\vskip0.5truecm
\addtocounter{figure}{1}
\centerline{Figure~\thefigure {\it a}}
\else\vskip-1.5truecm\figcaption{\figcapxfinalo}\fi
\end{figure}

\clearpage
\begin{figure}
%\epsfxsize=12.0truecm
%\plotone{xtellcor_final_order4_O6.5v.ps}
\plotone{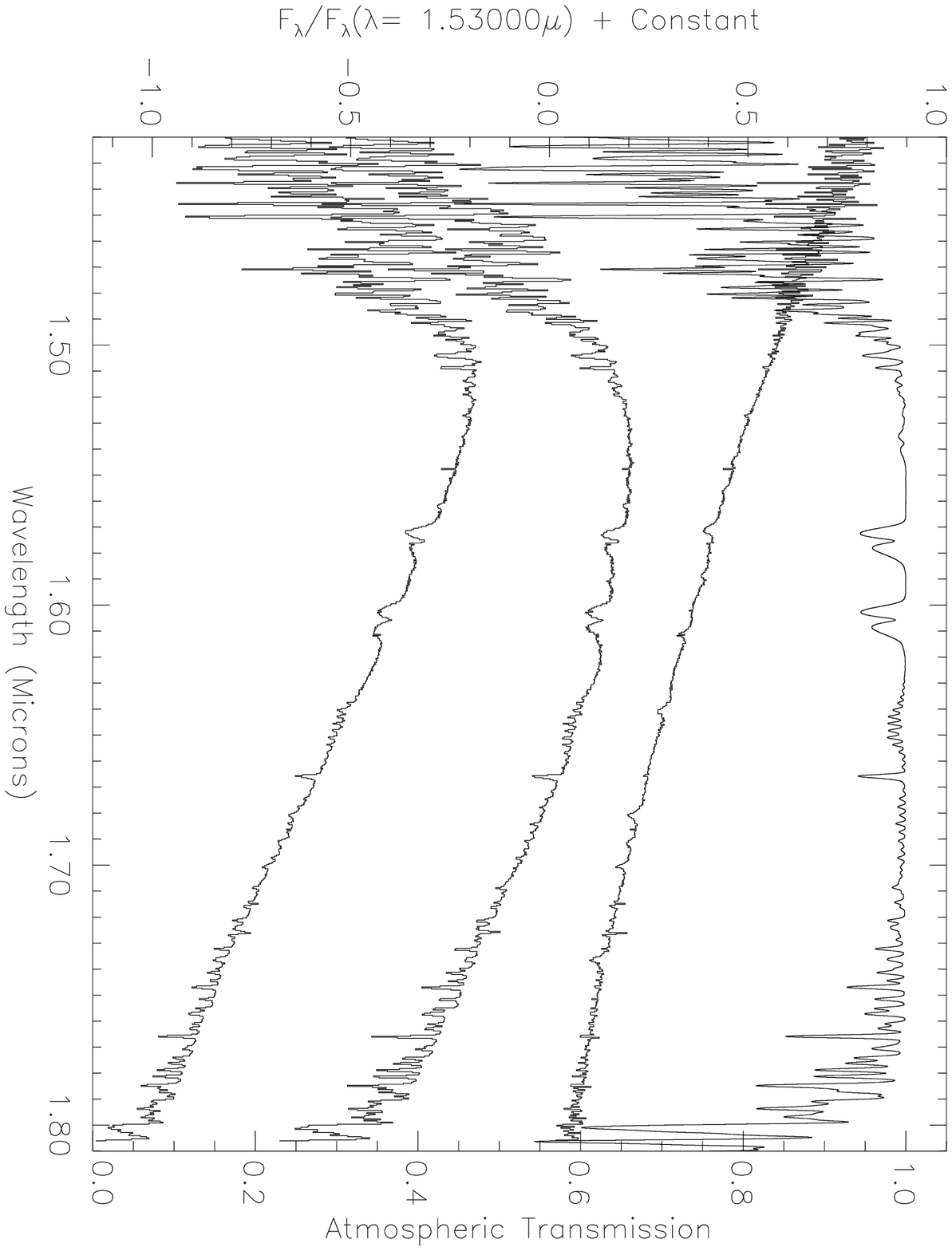}
\addtocounter{figure}{0}
\ifsubmode
\vskip0.5truecm
\else\vskip-0.3truecm\fi
\centerline{Figure~\thefigure {\it b}}
\end{figure}

\clearpage
\begin{figure}
%\epsfxsize=12.0truecm
%\plotone{xtellcor_final_order5_O6.5v.ps}
\plotone{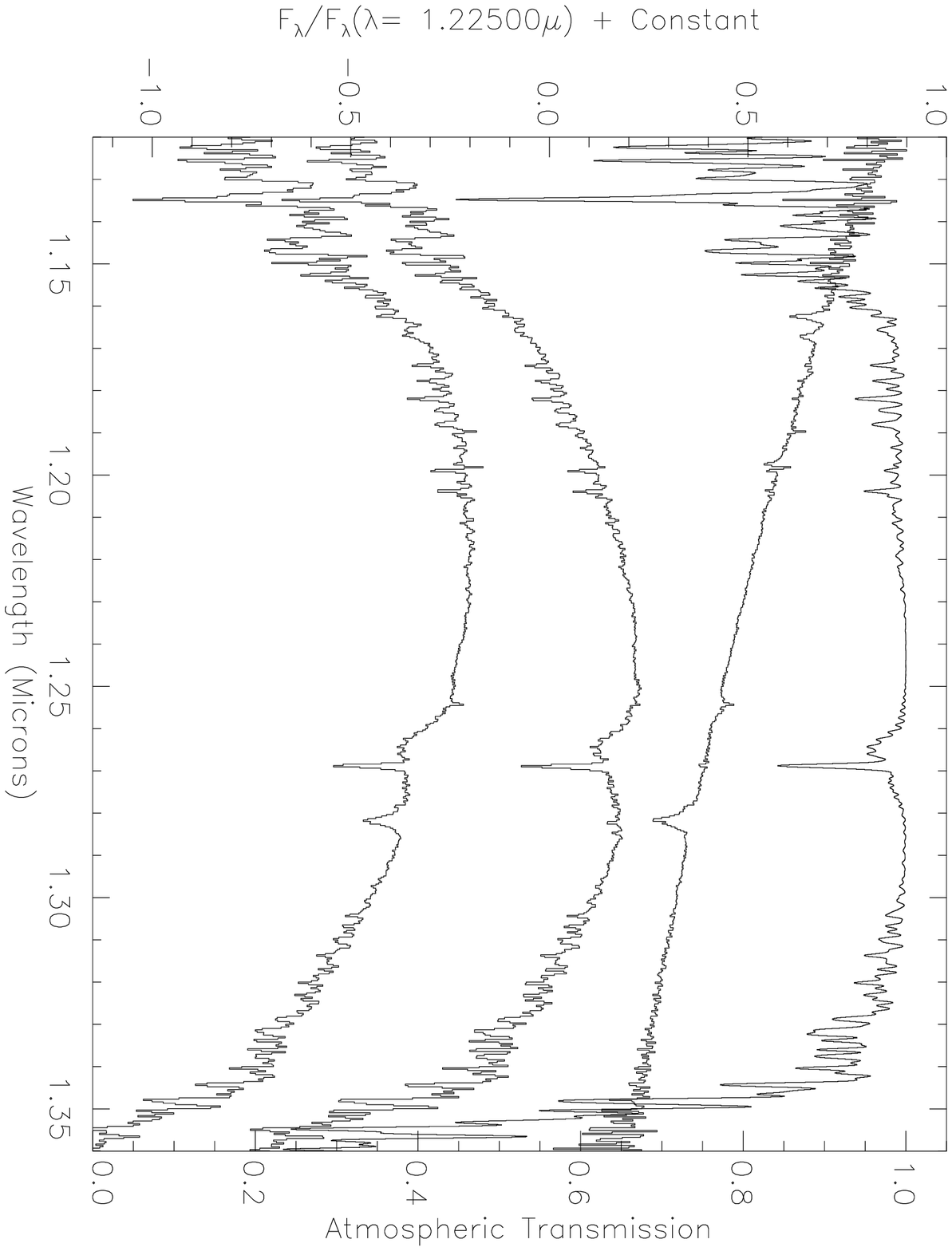}
\addtocounter{figure}{0}
\ifsubmode
\vskip0.5truecm
\else\vskip-0.3truecm\fi
\centerline{Figure~\thefigure {\it c}}
\end{figure}

\clearpage
\begin{figure}
%\epsfxsize=12.0truecm
%\plotone{xtellcor_final_order6_O6.5v.ps}
\plotone{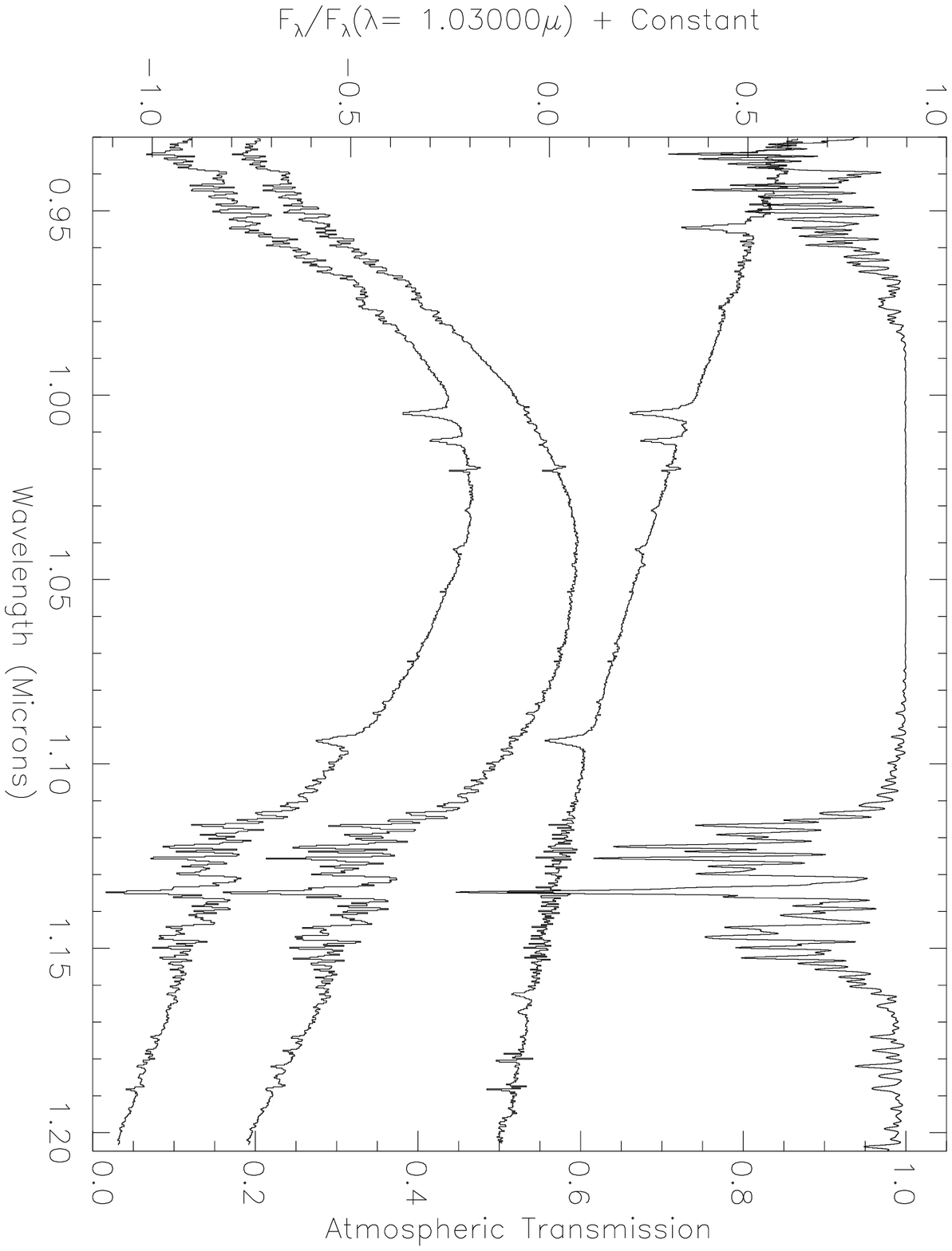}
\addtocounter{figure}{0}
\ifsubmode
\vskip0.5truecm
\else\vskip-0.3truecm\fi
\centerline{Figure~\thefigure {\it d}}
\end{figure}

\clearpage
\begin{figure}
%\epsfxsize=12.0truecm
%\plotone{xtellcor_final_order7_O6.5v.ps}
\plotone{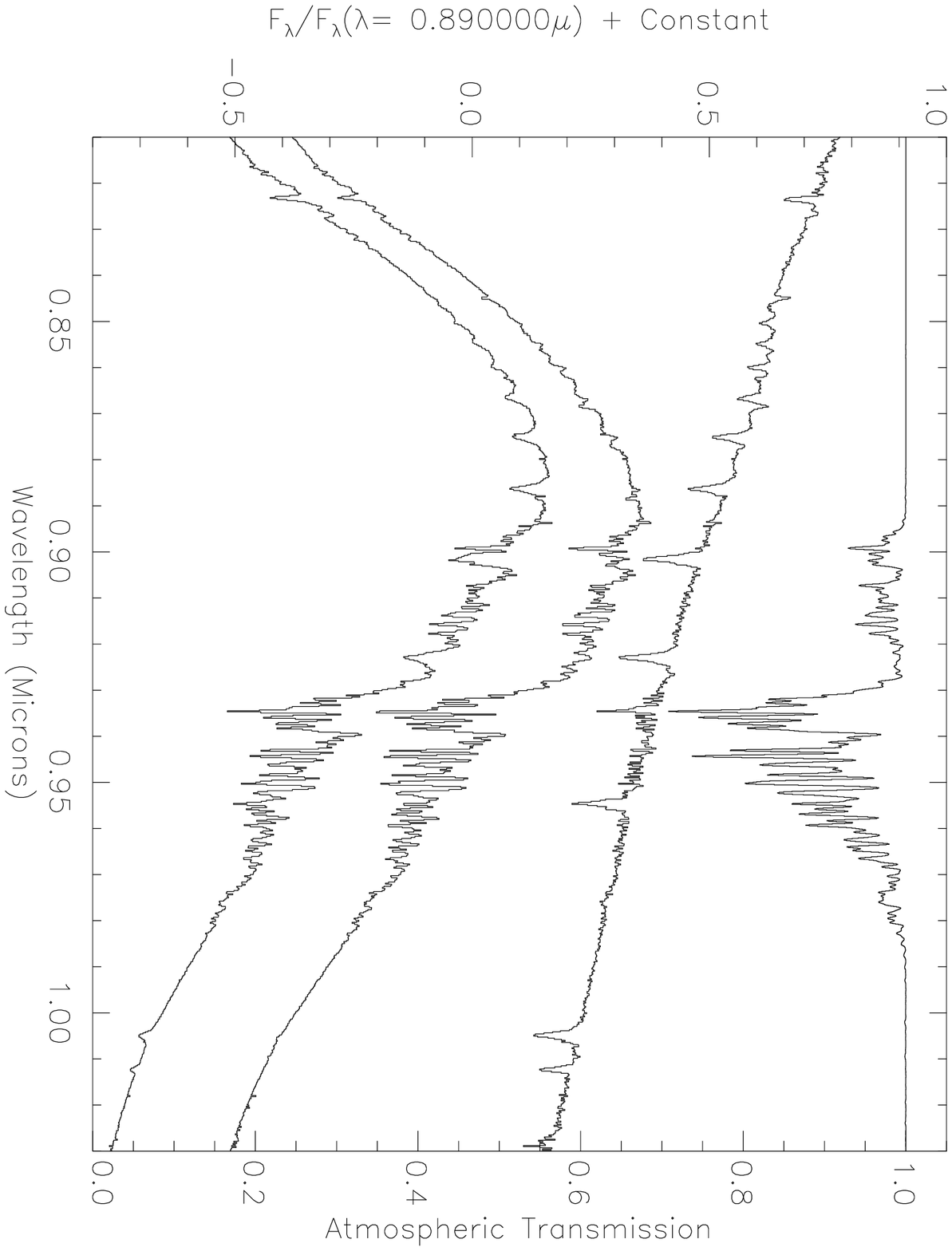}
\addtocounter{figure}{0}
\ifsubmode
\vskip0.5truecm
\else\vskip-0.3truecm\fi
\centerline{Figure~\thefigure {\it e}}
\end{figure}

\clearpage
\begin{figure}
%\epsfxsize=12.0truecm
%\plotone{xtellcor_final_order3_g8iii.ps}
\vskip-1.5truecm
\plotone{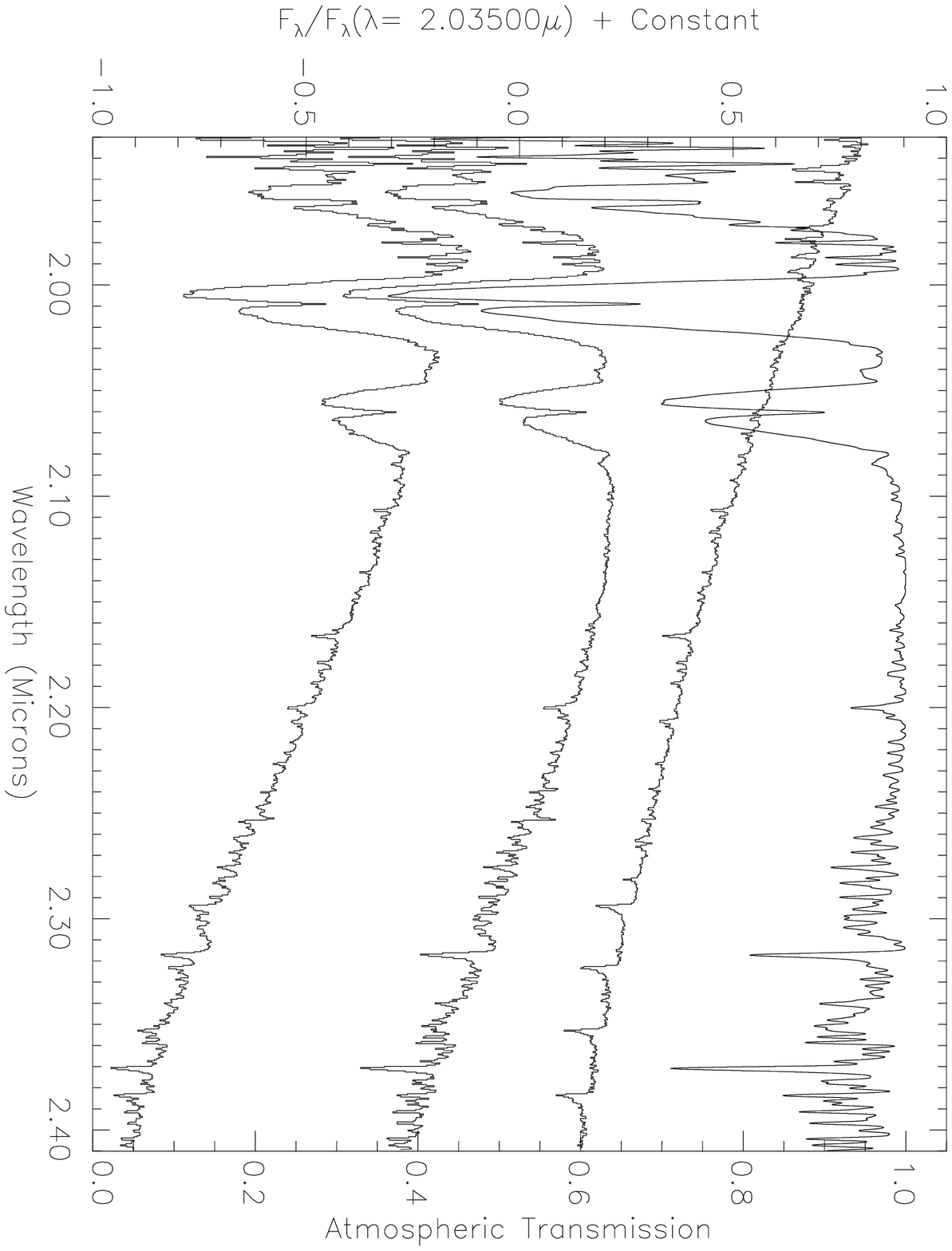}
\ifsubmode
\vskip0.5truecm
\addtocounter{figure}{1}
\centerline{Figure~\thefigure {\it a}}
\else\vskip-1.5truecm\figcaption{\figcapxfinalg}\fi
\end{figure}

\clearpage
\begin{figure}
%\epsfxsize=12.0truecm
%\plotone{xtellcor_final_order4_g8iii.ps}
\plotone{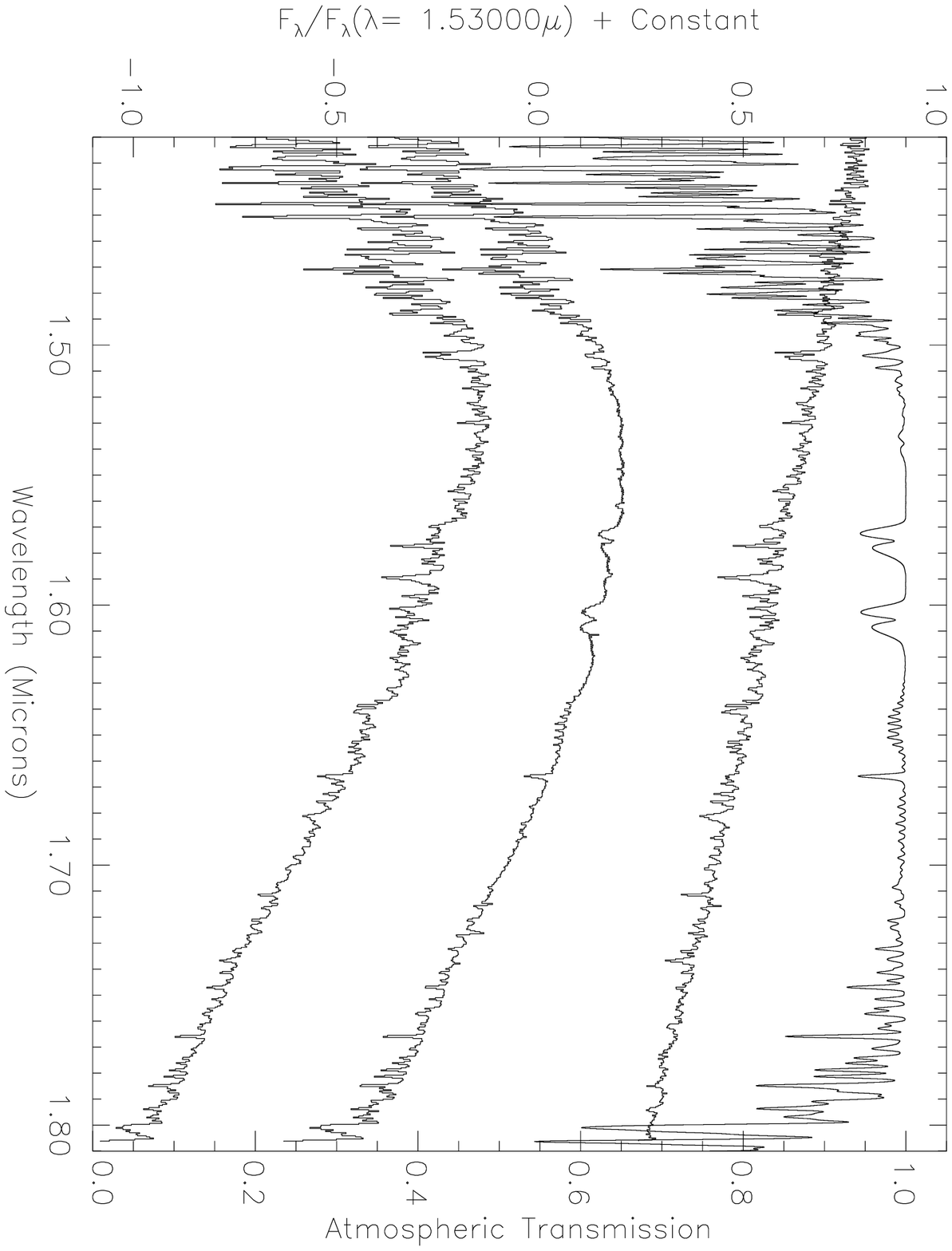}
\addtocounter{figure}{0}
\ifsubmode
\vskip0.5truecm
\else\vskip-0.3truecm\fi
\centerline{Figure~\thefigure {\it b}}
\end{figure}

\clearpage
\begin{figure}
%\epsfxsize=12.0truecm
%\plotone{xtellcor_final_order5_g8iii.ps}
\plotone{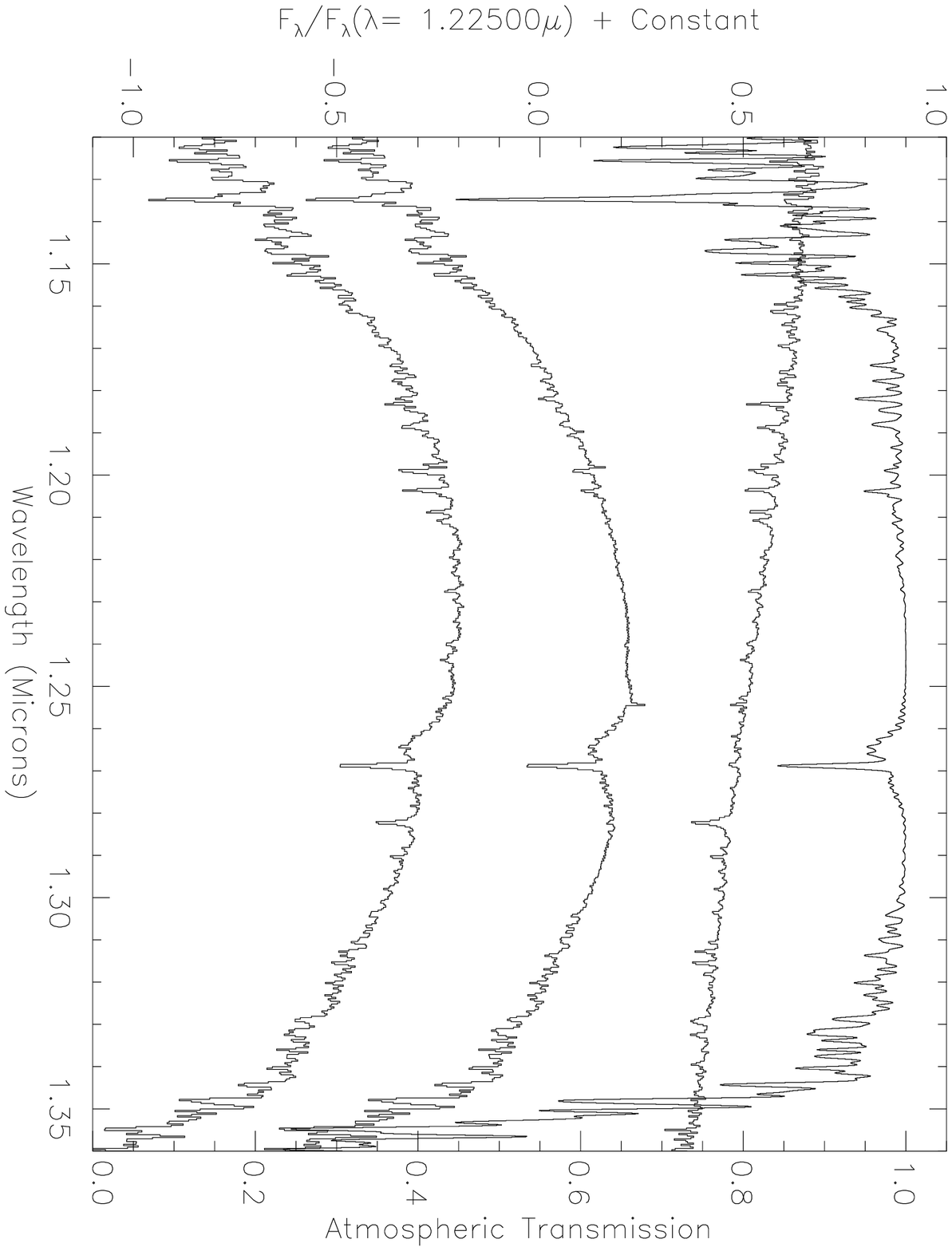}
\addtocounter{figure}{0}
\ifsubmode
\vskip0.5truecm
\else\vskip-0.3truecm\fi
\centerline{Figure~\thefigure {\it c}}
\end{figure}

\clearpage
\begin{figure}
%\epsfxsize=12.0truecm
%\plotone{xtellcor_final_order6_g8iii.ps}
\plotone{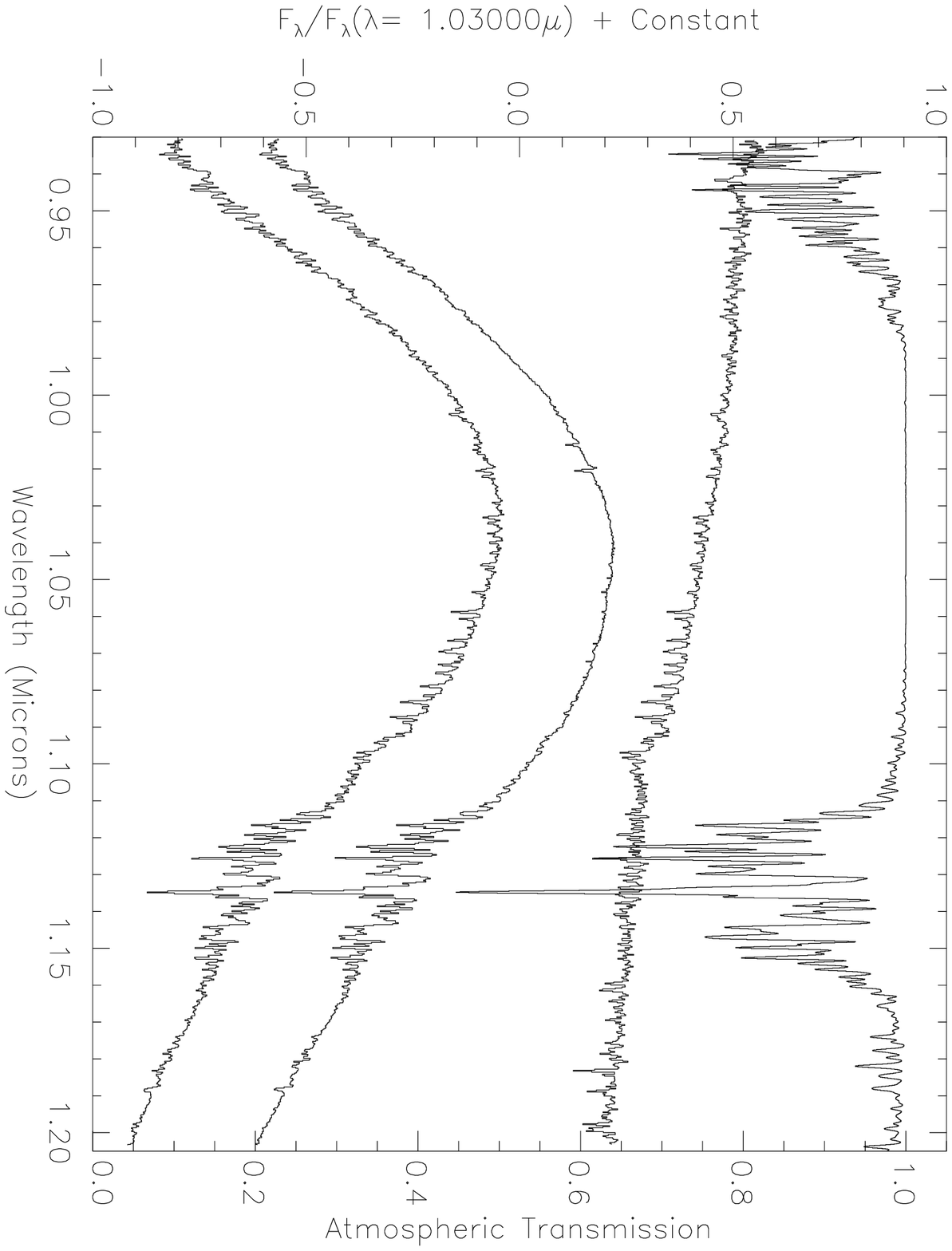}
\addtocounter{figure}{0}
\ifsubmode
\vskip0.5truecm
\else\vskip-0.3truecm\fi
\centerline{Figure~\thefigure {\it d}}
\end{figure}

\clearpage
\begin{figure}
%\epsfxsize=12.0truecm
%\plotone{xtellcor_final_order7_g8iii.ps}
\plotone{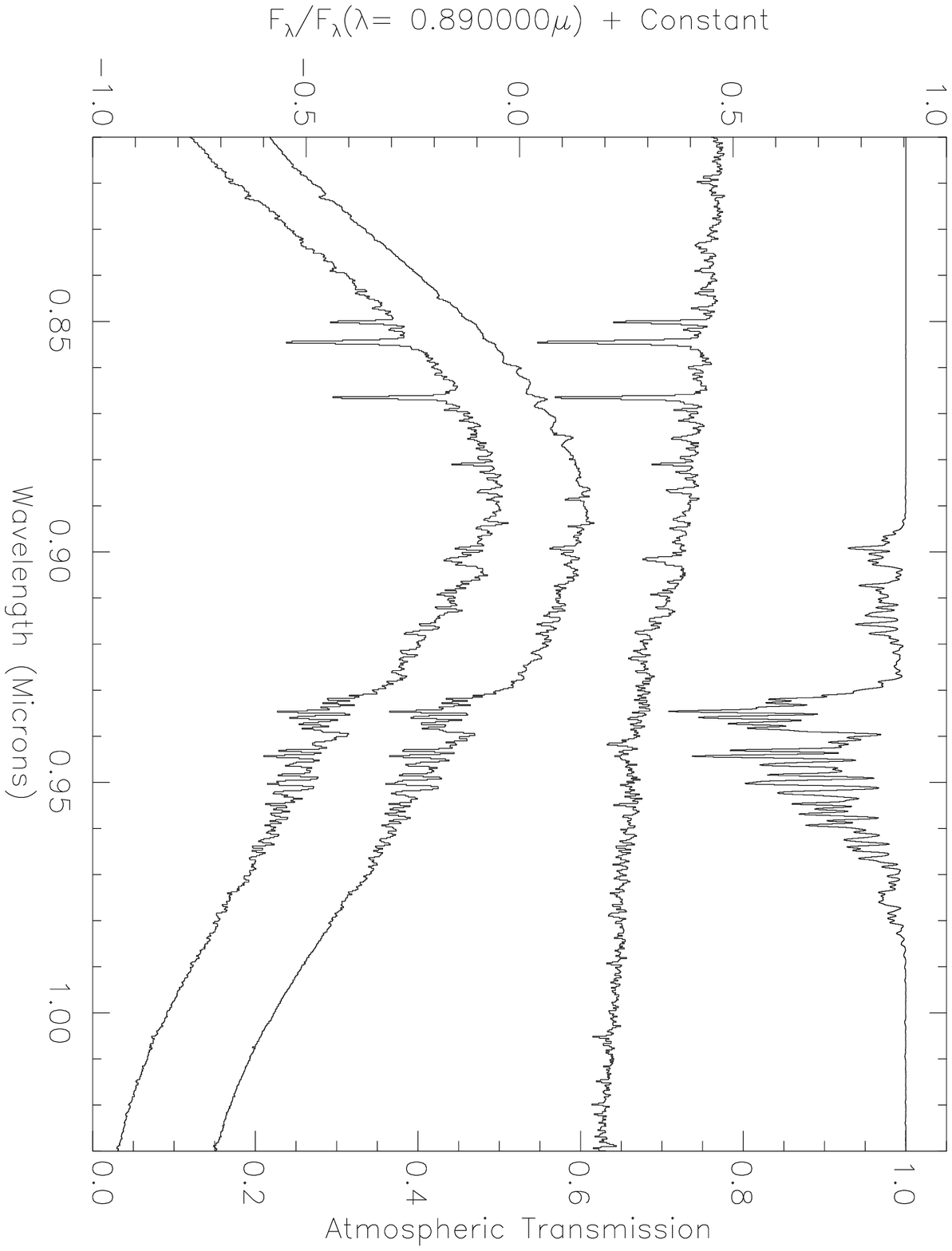}
\addtocounter{figure}{0}
\ifsubmode
\vskip0.5truecm
\else\vskip-0.3truecm\fi
\centerline{Figure~\thefigure {\it e}}
\end{figure}

%%% END OF FIGURES %%%

\fi

%%%%%%%%%%%%%%%
% End of Document
%%%%%%%%%%%%%%%
 
\end{document}